\documentclass[onecolumn]{emulateapj} 

\newcommand{\Teff}{T_{\mathrm{eff}}}
\newcommand{\Teffobs}{T_{\mathrm{eff,obs}}}
\newcommand{\Teffmod}{T_{\mathrm{eff,mod}}}
\newcommand{\FeH}{\left[\mathrm{Fe}/\mathrm{H}\right]}
\newcommand{\logg}{\log{g}}
\newcommand{\vsini}{v\sin{i}}
\newcommand{\obs}{{\mathrm{obs}}}
\newcommand{\mod}{{\mathrm{mod}}}
\newcommand{\dYdZ}{\frac{\delta Y}{\delta Z}}
\newcommand{\Mce}{M_{\mathrm{ce}}}
\newcommand{\Rce}{R_{\mathrm{ce}}}
\newcommand{\Msol}{M_\odot}
\newcommand{\Rsol}{R_\odot}
\newcommand{\MJ}{M_{\rm J}}
\newcommand{\RJ}{R_{\rm J}}
\newcommand{\Mpl}{M_{\rm pl}}
\newcommand{\Rpl}{R_{\rm pl}}
\newcommand{\Mstar}{M_\ast}
\newcommand{\Mearth}{M_\oplus}
\newcommand{\MV}{M_V}
\newcommand{\ks}{{\rm km/s}}
\newcommand{\be}{\begin{equation}}
\newcommand{\ee}{\end{equation}}
\newcommand{\bea}{\begin{eqnarray}}
\newcommand{\eea}{\end{eqnarray}}
\begin{document}
\submitted{To appear in \apjs}

\title{Structure and Evolution of Nearby Stars with Planets II.\\Physical Properties of $\sim1000$ Cool Stars from the SPOCS Catalog}

\shorttitle{Structure and Evolution of Nearby Stars with Planets II.}
\shortauthors{Takeda et al.}

\author{Genya Takeda\altaffilmark{1},
Eric B. Ford\altaffilmark{2},
Alison Sills\altaffilmark{3},
Frederic A. Rasio\altaffilmark{1},
Debra A. Fischer\altaffilmark{4} and
Jeff A. Valenti\altaffilmark{5} }

\altaffiltext{1}{Department of Physics and Astronomy,
Northwestern University, 2145 Sheridan Road, Evanston, IL 60208; {\tt g-takeda, rasio@northwestern.edu}}
\altaffiltext{2}{Department of Astronomy, University of California at Berkeley, 601 Campbell Hall, Berkeley, CA 94709; {\tt eford@astro.berkeley.edu}}
\altaffiltext{3}{Department of Physics and Astronomy, McMaster University, Hamilton, ON L8S 4M1, Canada; {\tt asills@mcmaster.ca}}
\altaffiltext{4}{Department of Physics and Astronomy, San Francisco State University, San Francisco, CA 94132; {\tt fischer@stars.sfsu.edu}}
\altaffiltext{5}{Space Telescope Science Institute, 3700 San Martin Drive, Baltimore, MD 21210; {\tt valenti@stsci.edu}}

\begin{abstract}
We derive detailed theoretical models for 1074 nearby stars from the SPOCS (Spectroscopic
Properties of Cool Stars) Catalog.   The California and Carnegie Planet Search has obtained
high-quality ($R\simeq 70000-90000$, $S/N\simeq 3-500$) echelle spectra of over 1000 nearby
stars taken with the Hamilton spectrograph at Lick Observatory, the HIRES spectrograph at
Keck, and UCLES at the Anglo Australian Observatory. A uniform analysis of the high-resolution
spectra has yielded precise stellar parameters ($\Teff$, $\log{g}$, $v \sin{i}$, [M/H] and
individual elemental abundances for Fe, Ni, Si, Na, and Ti), enabling systematic error
analyses and accurate theoretical stellar modeling.   We have created a large database of theoretical stellar evolution tracks
using the Yale Stellar Evolution Code (YREC) to match the observed parameters of
the SPOCS stars.   Our very dense grids of evolutionary tracks
eliminate the need for interpolation between stellar evolutionary tracks and allow precise determinations of physical stellar
parameters (mass, age, radius, size and mass of the convective zone, surface gravity, etc.).
   Combining our stellar models with the observed stellar atmospheric parameters and  uncertainties, we compute the likelihood for each set of stellar model parameters separated
by uniform time steps along the stellar evolutionary tracks.  The computed likelihoods are used for a Bayesian analysis to derive posterior probability
distribution functions for the physical stellar parameters of interest.  We provide a catalog of physical parameters for 1074  stars that are based on a uniform set of high quality spectral  observations, a uniform spectral reduction procedure, and a uniform set of stellar  evolutionary models.  We explore this catalog for various possible correlations between stellar and planetary properties,
which may help constrain the formation and dynamical histories of other planetary systems.
\end{abstract}


\section{Introduction}

	Precise analysis of high-resolution spectra of stellar atmosphere and theoretical calculations of the physical properties of low-mass stars are essential for a variety of astronomical problems.  Many previous works have focused on stellar abundance analyses and stellar age determinations to understand the chemical and dynamical history of the Galactic disk.  Since the benchmark work by \citet{edvardsson93a}, there have been a number of large observational surveys to obtain a true age -- metallicity relationship in the solar neighborhood \citep{feltzing01,ibukiyama02,nordstrom04}.  
	
	There has also been a recently growing interest in the  properties of stars with planetary companions.  Extensive Doppler radial-velocity surveys using high-resolution spectroscopy and large transit programs have made tremendous progress in the past decade and garnered nearly 200 extrasolar planets to date.  The distributions of observed planetary properties are of great importance for testing theories of planet formation and dynamics \citep{eggenberger04,marcy05b}.  The extensive spectroscopic observations in searches for planets have also revealed statistical differences between the planet-host stars and the normal dwarf stars with no companions.  It is now well established that the frequency of the giant planets with orbital period less than 3 years is a strong function of the stellar metallicity for solar-type stars.  Consequently, the planet-host stars exhibit a different metallicity distribution from that of single stars \citep{santos04, fischer05,santos05}.  The stellar atmosphere pollution by planet accretion has been proposed as one of the possible metal enrichment scenarios, and it has been tested by observations and theoretical models \citep{gonzalez97,laughlin97,sandquist98,pinsonneault01}.  The efficiency of metallicity enhancement by planet accretion is still under debate, but the models require an accurate knowledge of the mass of the stellar convective zone.  Deriving accurate distributions of stellar metallicity, age and convective zone size from a large sample of observations is useful for better understanding the origin of the metal-rich atmospheres of planet-host stars.

	Theoretical models of planet-host stars are also relevant to the dynamical studies of planetary systems.  A motivational work for this paper was done by \citet{ford99} (hereafter FRS99), who provided models for five stars with known ``hot Jupiters'' to estimate the orbital decay timescales for these systems.  Specifically,  they constrained the models of tidal dissipation and spin-orbit coupling in those systems using the estimated convective envelope masses and stellar ages.  Age estimates of planetary systems also help constraining long-term perturbations on planets.  For example, if a planet resides in a wide stellar binary system, the planetary orbit may undergo secular evolution, on a timescale as large as $\sim1\,$Gyr or even longer \citep{holman97,michtchenko04,takeda05,mudryk06}.   Estimates of the age and the physical properties of the host star can thus help constrain the dynamical history and the formation channel of the system.

	Despite the recent improvements in high-resolution spectroscopy and data analysis techniques, providing accurate stellar parameters is not a straightforward task, even for bright stars in the solar neighborhood.  Currently, there is no canonical method within data analysis or  theoretical modeling to estimate the stellar properties.  Various approaches have been applied, particularly for stellar age determinations.  The observed stellar rotation may be a good age indicator, since stars normally slow down as they age.   Chromospheric activities measured from the CA~II~H and K absorption lines are the favored rotation measure and therefore often used as a stellar age indicator \citep{wilson70,baliunas95,baliunas97,henry97,rocha98,henry00b}.  The stellar age can also be constrained by the surface lithium depletion \citep{soderblom83,boesgaard91}, though it needs to be treated with extra caution for planet-host stars as close-in planets may tidally affect the stellar convective envelope and thus cause further lithium depletion \citep{israelian04}.  The merits and challenges of different techniques have been well summarized by \citet{saffe05}.   They have carefully derived the ages of ~50 extrasolar planet host stars observed from the southern hemisphere and tested several different age determination schemes for comparisons.
	
	In this paper, we derive various stellar properties by matching the spectroscopically determined surface parameters to theoretical stellar evolution models.  This approach is similar to the traditional isochrone method.  In the isochrone method,  the observed $\MV$ and $B-V$ are placed  in the HR diagram, then  the stellar ages are derived by interpolating the observed position between the theoretically computed isochrones \citep{twarog80,vandenberg85,edvardsson93a,bertelli94,ng98,lachaume99,ibukiyama02,nordstrom04,pont04,jorgensen05,karatas05}.  Equivalently, stellar ages can be  derived by interpolating the observations between theoretically computed stellar evolutionary tracks.  Isochrone or evolutionary track analysis is becoming increasingly accurate relative to other methods such as age -- activity relations or age -- abundance relations, given the availability of advanced high-resolution echelle spectrographs,  sophisticated stellar evolutionary codes, and increased computational power.  Theoretical evolutionary models also have an advantage in that one can create a full model of a star, providing not just the stellar age but all the physical parameters, including those characterizing the stellar interior.

Note that  the traditional spectral data analysis using theoretical stellar evolutionary tracks (or equivalently, isochrones) involves many sources of systematic bias.  Here we summarize the main steps in the theoretical modeling together with the necessary precautions:

(i) {\it The observed sample} --- A substantial number of sample stars are required to obtain meaningful stellar parameter distributions.  The benchmark work by \citet{edvardsson93a} provided chemical abundances of 189 nearby field dwarfs.  Today, typical solar neighborhood surveys include $\sim$1000 to $\sim$10000 stars \citep{ibukiyama02,nordstrom04}.  As the sample size becomes less of a problem, however, the sample selection remains crucial for understanding the global statistics of the stellar properties.  Unfortunately, any survey is limited by its own selection criteria, and  in practice any sample  has some selection  effects.   The resultant observed statistics need to be analyzed very carefully to separate the selection effects from the true stellar properties.  \citet{nordstrom04} have done careful studies on the completeness of their sample stars, in terms of binarity, magnitude, sampling volume and other stellar parameters.  

	For this work, we have used the spectroscopic data from the SPOCS (Spectroscopic Properties of Cool Stars) catalog by \citet[][hereafter VF05]{valenti05}.  SPOCS consists of high-resolution echelle spectra of over 1000 nearby F-, G- and K-type stars obtained through the Keck, Lick and Anglo-Australian Telescope (AAT) planet search programs, including the 99 stars with known planetary companions.  The sampling criteria for the SPOCS catalog are such that the achievable Doppler velocity precision is optimized for planet detections.  Thus, the catalog favors stars that are bright, chromospherically inactive (or slow-rotating) on  the main-sequence or subgiant branch.  A subset of the catalog can be used for completeness studies in terms of sampling volume or the presence of planetary companions.

(ii) {\it Spectral data analysis} --- Currently there is no standardized technique for analyzing high-resolution spectra.  In the traditional abundance analysis approach, different radiative transfer and stellar atmosphere models produce different types of errors.  Apart from the choice of model, micro-turbulence needs to be carefully adjusted to match the equivalent width of the spectrum, and also macro-turbulence, stellar rotation, and instrumental profile need to be taken into account for the spectral line broadening effects.  All these factors may induce systematic biases in the derived model parameters.  

	In contrast to the traditional abundance analysis method, VF05 directly fit the observed spectrum to the synthetic spectrum generated by a software package SME \citep[Spectroscopy Made Easy,][]{valenti96}, allowing for a self-consistent error analysis.    Using SME, VF05 have derived a set of observational stellar parameters $m_V, T_{\rm eff}$, [Fe/H] and $\log{g}$ for each star, along with precise error estimates.  The spectral analysis techniques adopted for the SPOCS catalog are summarized in $\S\,$\ref{observation}.  For more details on the observations and SME pipelines, see \citet{valenti05}.

(iii) {\it Model parameter estimate} --- \citet{pont04} have thoroughly analyzed the systematic biases induced by the traditional maximum likelihood approach using theoretical isochrones, which simply interpolates the nearest isochrones to the observational data point in the HR diagram.  The simple isochrone interpolation approach accounts for neither  the highly non-linear mapping of time onto the  HR diagram nor the non-uniform mass and metallicity distributions of the stars in the galactic disk.  Consequently, the derived age distribution is biased toward an older age compared to the real distribution.  To avoid this bias, one needs to account for  the {\em a priori} distribution functions of stellar parameters.   For instance, the longer main-sequence timescale of lower-mass stars results in a smaller likelihood of observing low-mass post-main-sequence stars relative to higher  mass stars.   Bayesian probability theory including physically motivated prior distribution functions has been demonstrated to be an effective means of determining unbiased ages of stars in the solar neighborhood \citep{ng98,lachaume99,jorgensen05}. 

	To model the SPOCS stars, we have constructed large and fine grids of theoretical stellar evolutionary tracks, computed with the Yale stellar evolution code (YREC).  With YREC, the physical structure of the star can be calculated for each snapshot in time. Our procedure for evolutionary track calculations with YREC are described in detail in $\S\,$\ref{yrec}.  Tracks have been computed for more than 250000 stars with slightly different initial conditions in mass, metallicity and helium abundances (see $\S\,$\ref{grids} for the full description of the grids of stellar tracks).  The high resolution of the grids in time ($1\,$Myr) can apply accurate weighting for the accelerating evolutionary phases from the main sequence to post-main sequence.  Each of the stellar models from  each  of the stellar evolutionary tracks is then assigned to a cell in a four dimensional grid of observable stellar parameters to increase the computational efficiency.  Subsequently, {\em posterior} probability distribution functions (PDFs) are derived for each stellar parameter, in the framework of the Bayesian probability theory account ($\S\,$\ref{bayesian}).  

(iv) {\it PDF error estimate} --- A calculated PDF is normally summarized by a single value for the best estimate and/or a credible interval.  However, the choice of the best-fit value and the associated credible interval is often rather arbitrary.  There are at least three choices of statistical values commonly used to summarize the PDF: the mean, median and mode.  Since  derived PDFs are not necessarily  unimodal, let alone  normal distributions,   a single summary statistic will not  always be sufficient to accurately describe the posterior PDF.    Thorough analyses of accurate parameter representations have been done by \citet{jorgensen05} with extensive Monte Carlo simulations.  They have introduced the notion of ``well-defined'' age to distinguish the derived parameters whose posterior PDF have a well-defined peak within the given parameter range.   The same notation is adopted by \citet{nordstrom04} to derive the ages for more than 10000 stars from the Geneva-Copenhagen survey.  We have carefully presented all the necessary information from the calculated PDFs, applying a similar notation to that by \citet{jorgensen05}.  We will discuss the details of parameter estimates in $\S\,$\ref{bestestimate}.

	This paper is organized as follows.  In \S~2, we briefly describe the stellar sample, observations, data reduction and spectrum synthesis procedures.  In \S~3, we describe  the stellar  evolution code (\S~\ref{yrec}),  construction of grids of stellar evolutionary tracks (\S~\ref{grids}), the calculation of derived stellar parameters in the Bayesian framework (\S~\ref{bayesian}) and the error analysis for these  parameters (\S~\ref{pdfanalysis}).  The derived stellar properties and the parameter distributions for the SPOCS catalog stars are presented in \S~4.  The newly determined stellar properties of five planet-host stars previously modeled by FRS99 are presented in \S~\ref{comparison}.  In \S~\ref{additional_models}, we select five planet-host stars with particularly interesting stellar or planetary properties and discuss the derived models for these stars.  The overall stellar parameter distributions are further analyzed in \S~\ref{correlations}, for parameter correlations and constraints on various dynamical formation scenarios for extrasolar planets.

\section{The Spectral Analysis Technique \label{observation}}

VF05 have carried out 
a uniform spectroscopic analysis of 1944 spectra for a sample 
of 1140 FGK stars in planet search programs at Keck Observatory,
Lick Observatory and the AAT.
The stars for these surveys are selected to optimize the achievable Doppler 
velocity precision and favor bright, chromospherically inactive, main-sequence or subgiant 
stars ($M_V > 3.0$, $V < 8.5$ and $B-V > 0.5$).   Known stellar binaries with 
separations less than 2 arcseconds were rejected because the presence 
of a close stellar companion complicates the Doppler analysis. 

A detailed description of the methodology and an assessment of random and 
systematic errors is provided in VF05.  The spectral synthesis modeling 
program, SME \citep[VF05, ][]{valenti96} assumes 
local thermodynamic equilibrium and drives a radiative transfer code using Kurucz stellar 
atmosphere models \citep{kurucz93}, and atomic line data \citep[Vienna Atomic Line Database, 
VALD,][]{kupka99,ryab99} to create a synthetic spectrum. 
The code employs a non-linear least squares Marquardt fitting algorithm to 
vary free parameters ($\Teff$, $\logg$, $\vsini$ and abundances) in order to 
best match continuum and spectral line profiles in selected wavelength 
regions of an observed spectrum.  With each Marquardt iteration, the SME program 
interpolates over a grid of 8000 Kurucz stellar atmosphere models
before generating a new synthetic spectrum. Most of the SME analysis 
was made for wavelengths between $6000 - 6200 $ \AA\, to minimize problems 
with line blending.  An additional wavelength segment $5175 - 5190$ \AA\, 
was included to leverage the gravity sensitivity of the MgI b triplet lines.  

As discussed in VF05, the $1 \sigma$ uncertainties in the spectral 
modeling correspond to about $\pm 44\,$K for $\Teff$, $\pm 0.06\,$ dex for 
$\logg$, $\pm 0.03\,$dex for abundances, and $\pm 0.5 \, \ks\,$ for $v \sin i$.  
A number of researchers have carried out spectroscopic analyses for smaller subsets 
of stars in common with the VF05 sample. Effective temperatures and abundances show excellent 
agreement with these published results, however, small systematic offsets 
are seen. Comparisons of $\logg$ values show larger RMS scatter, but 
generally agree within quoted ($\sim0.15$ dex) uncertainties.  
While different spectroscopic analyses show reasonable agreement,
the non-negligible systematic offsets demonstrate that spectroscopic 
results from different investigators should be combined
with caution, particularly when looking for subtle correlations or trends. 
This large homogeneously analyzed sample is ideal for comparisons to the 
stellar evolutionary models here.

\section{Methods \label{methods}}

\subsection{Yale Stellar Evolution Code \label{yrec}}

We use the Yale Rotational Evolution Code (YREC) in its non-rotating
mode to calculate stellar models. YREC is a Henyey code which solves
the equations of stellar structure in one dimension
\citep{guenther92}. The chemical composition of each shell is updated
separately using the nuclear reaction rates of \cite{gruzinov98}. The
initial chemical mixture is the solar mixture of \cite{grevesse93}, scaled
to match the metallicity of the star being modeled. Gravitational
settling of helium and heavy elements is not included in these
models. For regions of the star which are hotter than $\log T (K) \geq
6$, we use the OPAL equation of state \citep{rogers96}. For regions
where $\log T (K) \leq 5.5$, we use the equation of state from
\cite{saumon95}, which calculates particle densities for hydrogen and
helium including partial dissociation and ionization by both pressure
and temperature.  In the transition region between these two
temperatures, both formulations are weighted with a ramp function and
averaged.  The equation of state includes both radiation pressure and
electron degeneracy pressure.  We use the OPAL opacities \citep{iglesias96}
for the interior of the star down to temperatures of $\log T (K) =
4$. For lower temperatures, we use the low-temperature opacities of
\cite{alexander94}. For the surface boundary condition, we use the stellar
atmosphere models of \cite{allard95}, which include molecular effects. We
use the standard B\"{o}hm-Vitense mixing length theory
\citep{cox68,BV58} with $\alpha$=1.7161. This value of $\alpha$, as
well as the solar hydrogen abundance, $X_{\odot}=0.70785$, is 
obtained by calibrating models against
observations of the solar radius ($6.9598 \times 10^{10}$ cm) and
luminosity ($3.8515 \times 10^{33}$ erg/s) at the present age of the
Sun ($4.57\,$Gyr). 

\begin{figure}
  \begin{center}
    \includegraphics[width=0.6\columnwidth]{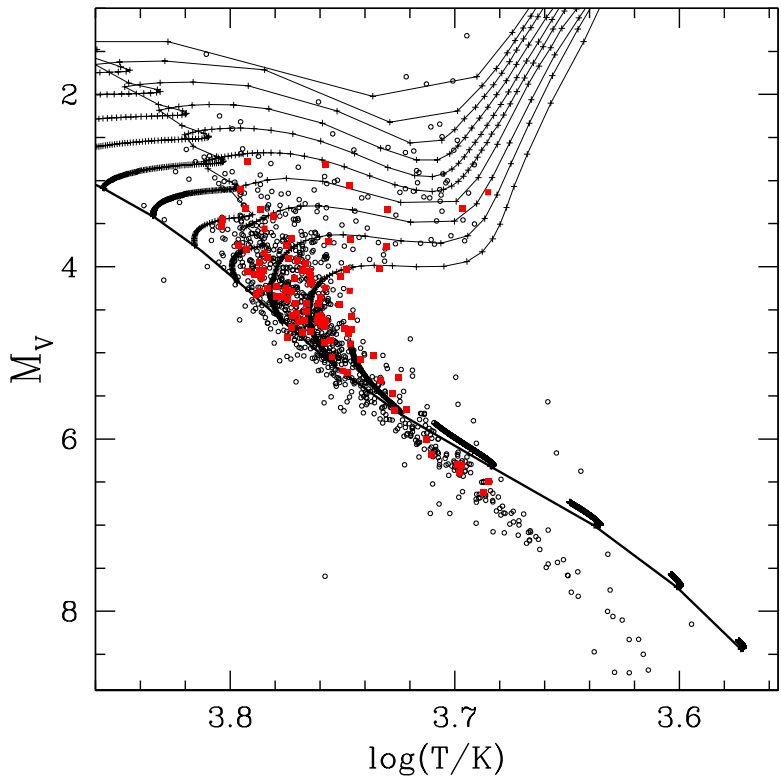}
    \caption{Sample stellar evolutionary tracks with the solar metallicity ($X \approx 0.71, Z \approx 0.02$) computed with YREC.   Here 99 stars with known planetary companions (red solid circles) and 975 stars with no detected planetary companions (open circles) are overlaid.  The thin solid lines with $\times$ are the theoretical evolutionary tracks of stars for every $0.1 M_\odot$ from 0.5 to 2.0 $M_\odot$, evolved up to 14\,Gyr.  The evolutionary tracks for stars with $M < 1.0\,\Msol$ reach 14\,Gyr during the main-sequence phase.    Each $\times$ on the tracks is separated by 20\,Myr.  Note that the evolutionary sequence accelerates in the post-main-sequence evolution, corresponding to the Hertzsprung gap.  The thick solid line represents the location of the ZAMS stars.  The diagram represents only a small part of our grid of evolutionary tracks, which encompasses much wider ranges of $\Teff, L, \FeH$, densely covered by the tracks that evolved from different initial conditions. 
      \label{samplegrid.fig}}
  \end{center}
\end{figure}

\subsection{Constructing the Grids of Stellar Evolutionary Tracks \label{grids}}

Our grids consist of 246000 theoretical stellar evolutionary tracks, separated by small intervals of mass and metallicity.  In this section, we describe the iteration method we have adopted to construct these grids of model tracks.  

First, we have constructed models of pre-main-sequence (PMS) stars with
masses ranging from 0.5 to 2.0 $M_\odot$ for every 0.1 $M_\odot$.
These PMS stars are modeled by solving the Lane-Emden equation for a
polytrope of index n = 1.5.  Each star in the grids has evolved from 
a polytrope with the mass closest to the desired value.  Since our
grids have much finer mass bins ($dM = 0.001 M_\odot$) and a wide range 
of metallicity (from -1.0 to +0.60 dex in [Fe/H]), 
we first rescaled the selected polytrope to the desired values of
mass and composition (X,Y and Z), then numerically relaxed it.
After these parameters are rescaled, the star begins its PMS evolution
through the zero-age main sequence (ZAMS) into the main sequence (MS) and
subsequent evolutionary stage. 
The tracks through the PMS phase up to the ZAMS are
not included in our modeling, since the SPOCS catalog would reject such young stars on the basis of strong Ca H and K emission

Using YREC we calculate all the model parameters of interest 
($L, T_{\rm eff}$, $\log{g}, R$ and so on) for each
snapshot of the evolutionary sequence, starting from the ZAMS, 
until either (i) $14\,$Gyr is reached, or (ii) the end of the 
subgiant branch (beginning of the red-giant phase), whichever comes first.
Table~\ref{resolution.tab} describes the sets of initial stellar
parameters from which stars are evolved.  We vary the mass from 0.5
to 2.0 $M_\odot$ for every 0.001 $M_\odot$, and [Fe/H] from -1.0 to
0.6 dex for every 0.04 dex.  YREC takes the compositional mass fractions
$\{X, Y, Z\}$ as input, thus the iron abundance is converted into 
the mass ratio of heavier elements, Z, as $Z = Z_\odot \times 10^{[{\rm Fe/H}]}$,
using the assumed solar value $Z_\odot =$ 0.0188 \citep{grevesse93}.  For the helium
abundance Y, the previous work by \citet{ford99} adopted the canonical 
linear relation between the abundances of helium and heavier elements
\citep{pagel92,bressan94,edvardsson93a,edvardsson93b},
\begin{equation}
\frac{\delta Y}{\delta Z} \equiv \frac{Y - Y_\odot}{Z - Z_\odot},
\end{equation}
applying the solar helium abundance of $Y_\odot$ = 0.27335.  We have also adopted the linear correlation between Y and Z, varying the values to be $\delta Y / \delta Z$ = 0.0, 1.5, 2.5 and 3.5, assuming a  uniform  prior distribution of $\delta Y / \delta Z$.

YREC solves a set of linearized equations using adaptive time steps
$\delta \tau$ which are consecutively determined from a number of
criteria.  While YREC uses adaptive time steps to evolve the stellar models, we have interpolated 
all the evolutionary tracks to uniform time steps so as to properly account for the likelihood of observing a star at a given point in its evolution.  
Specifically, we have adopted $\Delta \tau$ = 20 Myr
for the stars with $M < 1.0 M_\odot$ and $\Delta \tau$ = 1 Myr for the
stars with $1.0 M_\odot \le M \le 2.0 M_\odot$.  The finer time resolution
for the stars with higher mass is justified as follows.  In 
the HR diagram, an evolutionary track of a star with a mass larger than  $\sim1.2 M_\odot$ has 
a sharp hook at the turnoff from the main-sequence track when the convective core
contracts sharply due to the flat hydrogen profile at core hydrogen
exhaustion (see Figure~\ref{samplegrid.fig}). To account for this sharply non-monotonic behavior, a
greater resolution of time is required for the stars with larger
masses.  When computing posterior PDFs, we weight each model by inverse of the time step, so as to account for  the difference in the interpolated time step and to maintain the uniform prior distribution in age (see \S~\ref{bayesian}).  

\begin{deluxetable}{lrr}
\tabletypesize{\scriptsize}
\tablecaption{Resolution of the stellar model grid \label{resolution.tab}}
\tablewidth{0pt}
\tablehead{
	\colhead{X} & \colhead{range} & \colhead{$\Delta X$}
}
\startdata
Mass 		& 0.5 -- 2.0 $M_\odot$ 	& 0.001 $M_\odot$ \\
${\rm [{\rm Fe/H}]}$ 	& -1.0 -- 0.6 dex 				& 0.04 dex \\
dY/dZ 	& 0.0, 1.5, 2.5, 3.5		&\\
age 		& 0.0 -- 14.0 Gyr 				& 20 Myr (M $< 1.0M_\odot$) \\
				& 											& 1 Myr (M $\geq 1.0 M_\odot$)\\

\enddata
\end{deluxetable}

\subsection{Bayesian Analysis \label{bayesian}}

To explore quantitatively the observational constraints on the stellar
parameters, we employ the techniques of Bayesian inference.  Our
approach closely follows that of \citet{pont04}, but we have
generalized their methods as described below.  Most notably, we:
1) include observational measurements of the surface gravity, 2) take
into account the correlations between the derived values of the
stellar metal abundance, surface gravity, and effective temperature,
and 3) compute a dense set of stellar models that eliminates the need
for interpolation between stellar tracks.

In the Bayesian framework, the model parameters are
treated as random variables which can be constrained by the actual
observations.  Therefore, to perform a Bayesian analysis, it is
necessary to specify both the likelihood (the probability of making a
certain observation given a particular set of model parameters) and
the prior (the {\em a priori} probability distribution for the model
parameters).  Let us denote the model parameters by $\theta$ and the
actual observational data by $d$, so that the joint probability
distribution for the observational data and the model parameters is
given by
\be
p(d, \theta) = p(\theta) p(d | \theta) = p(d) p(\theta | d), 
\ee
where we have expanded the joint probability distribution in two ways
and both are expressed as the product of a conditional  probability distribution
($p(d|\theta)$ or $p(\theta|d)$) and a marginalized probability distribution ($p(\theta)$ or $p(d)$).
A marginalized probability distribution can be calculated by simply integrating  
the joint probability distribution over all but one  of  the variables (e.g.,
$p(\theta) \equiv \int d\theta p(d,\theta)$).

Often the model parameters contain one or more parameters of particular interest (e.g.,
the stellar mass, radius, and age) and other ``nuisance parameters'' that are necessary
to adequately describe the observations but that we are not particularly interested in 
measuring for their own sake (e.g., the distance, helium abundance).

We assume that the function mapping the  stellar models to the observational data 
is a surjection or ``onto  function'', meaning that we assume that  our domain of  
stellar models includes at least one model that would result in any possible set
of true  values of the observable stellar parameters.  We use Bayes' theorem to 
calculate posterior probability distributions for the model parameters, as well as
other  physical quantities derived from the stellar models.  

Ideally, we would employ a hierarchical Bayesian analysis that could also incorporate the uncertainties in the theoretical models (e.g., choice of equation of state, opacity tables, treatment of convection, etc.), so as to estimate the uncertainties in physical parameters even more accurately.  Unfortunately, this is not yet computationally practical.  While we have attempted to minimize such complications (at least for comparisons between stars in our sample) by making use of a single set of state-of-the art stellar models and the largest available uniform set of high quality spectroscopic observations and determinations of stellar atmospheric parameters, we acknowledge that our uncertainty estimates may not fully account for the systematic difference between different stellar evolutionary codes.

\subsubsection{Priors}

We assume a prior of the form $p(M,\tau,D,X,Y,Z) = p(M) p(\tau) p(D)
p(X,Y,Z)$.  Thus, the priors for the mass, age, and distance are
independent of each other as well as the chemical composition.
However, the hydrogen, helium, and metal abundances are correlated.
If we write $p(X,Y,Z) = p(Z) p(Y|Z) p(X|Y,Z)$, then the logical
constraint that $X+Y+Z = 1$, implies that $p(X | Y,Z) =
\delta(1-Y-Z)$, where $\delta$ is the Dirac delta function.  Given the
observational difficulties in measuring $Y$ for main-sequences stars,
we base the helium abundance on the metallicity using an assumed value
of $\delta Y / \delta Z$.  Thus, we assume $p(Y|Z) = \delta(Y - Y_\odot
- (Z - Z_\odot) \delta Y/\delta Z)$.  While chemical abundance studies
of a variety of stellar populations suggest $\delta Y/\delta Z\simeq
2.5$ (see \S~\ref{grids}), we employ a hierarchical model in which we
assume a prior for $\delta Y/\delta Z \sim U[0,3.5]$.  We write the
prior for the metal fraction in terms of the surface metallicity, $Z =
Z_\odot 10^{\left(\FeH - \FeH_\odot\right)}$.

Since the SPOCS catalog is intentionally biased towards high
metallicities, we have chosen to adopt a uniform prior for the
metallicity, $p(\FeH) \sim U[-1.0, 0.6]$, instead of using an empirical 
distribution for stars in the solar neighborhood.  We implicitly assume that
the abundance of all metals is proportional to the iron abundance.
Thus, our prior can be written as $p(M,\tau,\FeH,D)$.

For the prior for stellar mass, we use a truncated power law based on
empirical estimates of the IMF, $p(M) \sim M^{-2.35}$ for $0.5 M_\odot = M_{\min} 
\le M \le M_{\max} = 2.0 M_\odot$.  Fortunately, the observations typically
provide a tight enough constraint on the stellar mass that the results
are relatively insensitive to the exact form of the mass prior.  Since
we impose sharp upper and lower cutoffs on the prior for the stellar
mass, we are able to compute an extremely dense grid of stellar
evolution tracks for relatively small range of stellar masses.  The
choice of the upper and lower limits is based on the selection
criteria for the VF05 sample that we analyze in
this paper.  While our models and methods can be applied to other
observations, the current set of stellar evolution tracks is limited
to stars that are almost certainly between $M_{\min}$ and $M_{\max}$.

We adopt a prior that is uniform in stellar age, $p(\tau) = 1/\tau_{\max}$
for $0$ yr $\le t \le t_{\max} = 14\,$Gyr.  This choice represents the maximum
entropy prior satisfying the obvious logical criterion that all stars
have a positive age that is less than the age of the universe from
cosmological observations \citep{spergel03}.  Since determining
stellar ages is notoriously difficult and the interpretation of the
ages of a population of stars is subtle \citep[e.g.,][]{pont04}, we
intentionally avoid incorporating prior observational or theoretical
notions about the star formation history of the solar neighborhood.
Indeed, we believe that this work has the potential to shed light on
the history of star formation in the solar neighborhood, based on the
combination of a large uniform sample of stellar atmospheric
parameters determined with high resolution spectroscopy, a large,
dense set of stellar evolutionary tracks, and our use of Bayesian
inference.  Note that this choice would be optimal for a stellar
population that had a constant rate of star formation over $\tau_{\max}$.
This is reassuring, since the observations are reasonably
approximated by a globally constant star formation rate for the
galactic disk.

For the prior distribution of stellar distances, we assume a uniform
density, $p(D) \sim D^2$, for $1$pc $= D_{\min} \le D \le D_{\max} =
10\,$kpc.  The lower (upper) limit is intentionally chosen to be
sufficiently small (large) that they are clearly excluded by the
Hipparcos parallax measurements for all the stars that we analyze in
this paper and hence do not affect the posterior probability
distributions.  The parallax is related to the distance by the
definition $\pi/\mathrm{arcsec} = \mathrm{pc}/D$.

Note that the above priors can be thought of as implicitly defining a
set of priors for observables (e.g., parallax, luminosity, effective
temperature, and surface gravity) via the mapping provided by the
stellar evolutionary models (see \S~\ref{yrec}).  We emphasize that a uniform
age prior maps into a very non-uniform distribution in the observable
quantities due to large changes in the time derivatives of the
observable quantities with stellar age.  Thus, we expect that our
proper Bayesian treatment may result in significantly different and
statistically superior estimates of physical quantities when compared
to traditional frequentist analyses.  Indeed, this is one of the
primary motivations for our reanalysis of the VF05
 sample.

\subsubsection{Likelihood}

We regard the observed parallax as being normally distributed about
the actual parallax with a dispersion given by the standard error
reported in the Hipparcos catalogue.  In practice, the uncertainty in
the stellar luminosity is dominated by the uncertainty in the parallax
measurement, so the visual magnitude and bolometric correction can be
regarded as exact.  Note that the combination of parallax, visual
magnitude, and bolometric correction place an observational constraint
on the stellar luminosity that is asymmetric and has a positive skewness.

We assume that the observational uncertainties in the stellar visual
magnitude and parallax are independent of each other and the derived
atmospheric parameters.  However, we do account for the significant
correlations between the derived effective temperature, chemical
composition, and surface gravity.  We assume that the stellar
atmospheric parameters derived from spectroscopic observations by
VF05 are distributed according to a multi-variate
normal distribution about their true values.  Unfortunately, adopting
a covariance matrix ($C$) based on the information matrix evaluated at
the best-fit set of stellar parameters leads to significantly
underestimating the observational errors, as determined by comparing
analyses of the same star using multiple spectroscopic observations.
This is due to the $\chi^2$ surface being very ``bumpy'' and the
curvature at any one local minima not reflecting the probability of
the true solution being in another local minima.  Since VF05
 have multiple observations of several stars in their
sample, they are able to compare the atmospheric parameters that they
derive from multiple observations of the same star to more accurately
determine their measurement errors (see VF05).  We
extend this approach to also determine the off-diagonal terms of the
covariance matrix based on 9 sets of the derived solar atmospheric
parameters (based on multiple observations of Vesta).  For each star,
we scale each of the empirically determined solar covariance terms by
the standard error of both the relevant parameters for the target
star.

Thus, our likelihood function is
\begin{eqnarray}
L(V, \pi, \log \Teff, \FeH, \log g ) & = & L_H(V, \pi, \log \Teff, \FeH, \log g ) \times L_S(V, \pi, \log \Teff, \FeH, \log g ) \nonumber \\
& = & \frac{\delta(V-V_\obs)}{(2\pi)^{3/2}\sigma_\pi } 
\times \exp\left[ -\frac{\left(\pi_\obs - 10^{-\left(V_\obs-M_{V,\mathrm{mod}}\right)/5+1} \right)^2}{2\sigma_\pi^2}\right]  \nonumber \\ 
& & \times \frac{1}{\sqrt{2\pi\mathrm{det}(C)}}\exp\left[ -\frac{1}{2} \left(d^S_\obs-d^S_\mod\right)^T C^{-1} \left(d^S_\obs-d^S_\mod\right) \right]
\end{eqnarray}
where $\sigma_\pi$ is the measurement uncertainty in the parallax,
$\mathrm{det}(C)$ is the determinant of the covariance matrix adopted
for $d^S_\obs$, $M_{V,\mathrm{mod}} =
M_{V,\mathrm{mod}}(M,\tau,\FeH,\delta Y/\delta Z)$ is the absolute
magnitude in the V band from the model (that includes a bolometric
correction), and $d^S_\mod = d^S_\mod(M,\tau,\FeH,\delta Y/\delta Z)$ is
the set of $\log \Teff$, $\FeH$, and $\log g$ for the model of a star
of mass $M$, metallicity $\FeH$, age $\tau$, and helium abundance implied
by $\delta Y/\delta Z$.  Note that we have separated the likelihood
into two components, one that is a function of $V_\obs$ and
$\pi_\obs$, and another that is a function of the spectroscopic
parameters, $d^S_\obs$.

\subsubsection{Posterior}

The posterior probability for a set of model parameters $(M,\tau,\FeH,D)$
given the observational data $d$ is given by
\begin{eqnarray}
& & p(M,\tau,\FeH,\delta Y/\delta Z,D | V_\obs, \pi_\obs, \log \Teffobs, \FeH_\obs, \log g_\obs) = \nonumber \\ 
& & \frac{ p(M,\tau,\FeH,\delta Y/\delta Z,D) L(V_\obs,\pi_\obs,\Teffobs,\FeH_\obs,\log g_\obs)}
{ \int dV\, d\pi\, d\Teff\, d\FeH\, d\log g \, p(M,\tau,\FeH,\delta Y/\delta Z,D) L(V,\pi,\Teff,\FeH,\log g) },
\end{eqnarray}
where the integral is formally over the entire range of possible
visual magnitudes, parallaxes, effective temperatures, metallicities,
and surface gravities.  When there are multiple spectroscopic
observations of the same star, then the spectroscopic portion of the
likelihood, $L_S$, is replaced by a product of multiple $L_S$'s, with
one $L_S$ being evaluated with each of the observed sets of
spectroscopic parameters.

Often, we are particularly interested in the posterior probability
density marginalized over all but one of the model parameters.  This
can be easily calculated from the posterior probability density by
integrating over all but one of the model parameters.  For example,
the marginal density for the stellar mass is given by
\begin{eqnarray}
& & p(M | V_\obs, \pi_\obs, \log \Teffobs, \FeH_\obs, \log g_\obs)  = \nonumber \\
& &  \int  d\tau\, d\FeH\, d\delta Y/\delta Z\, dD \, p(M,\tau,\FeH,\delta Y/\delta Z, D | V_\obs, \pi_\obs, \log \Teffobs, \FeH_\obs, \log g_\obs). 
\end{eqnarray}
We directly calculate the marginal posterior densities for the stellar
mass, age, and metallicity.  We also calculate marginalized posterior
densities for derived physical quantities.  For example, the marginal
posterior density for the stellar radius, $R$, is given by by
\begin{eqnarray}
& & p(R | V_\obs, \pi_\obs, \log \Teffobs, \FeH_\obs, \log g_\obs) =  \nonumber \\
& & \int dM\, d\tau\, d\FeH\, d\delta Y/\delta Z\, dD \, \delta(R-R_{\mathrm mod}(M,\tau,\FeH,\delta Y/\delta Z)) \nonumber \\
& & p(M,\tau,\FeH,\delta Y/\delta Z,D | V_\obs, \pi_\obs, \log \Teffobs, \FeH_\obs, \log g_\obs),  
\label{EqnMarginal}
\end{eqnarray}
where $R_{\mathrm mod}(M,\tau,\FeH,\delta Y/\delta Z)$ is the radius of
the stellar model with mass $M$, age $\tau$, metallicity $\FeH$, and
helium abundance implied by $\delta Y/\delta Z$.

\subsubsection{Numerical Methods}

The main difficulty in performing Bayesian analyses is the difficulty
of performing all the necessary integrals, particularly in high
dimensional parameters spaces.  Here, we describe the numerical
methods we use to approximate the necessary integrals.

To numerically compute these marginal densities, we discretize the
integrals in the standard way.  So Eqn.\ \ref{EqnMarginal} becomes
\begin{eqnarray}
&& p(R_o \le R \le R_o + \Delta R | V_\obs, \pi_\obs, \log \Teffobs, \FeH_\obs, \log g_\obs) \simeq  \nonumber \\
&& \sum_{i} p(M_i) \Delta M_i \sum_{j} p(\FeH_j) \Delta \FeH_j \sum_k p(t_k) \Delta t_k \sum_l p\left(\dYdZ_l\right) \Delta \dYdZ_l \times \nonumber \\
&& I\left[R_o \le R_{\mathrm mod}\left(M_i,\tau_k,\FeH_j,\dYdZ_l\right) \le R_o + \Delta R\right] \nonumber \\
&& 
\times p(M_i,\tau_k,\FeH_j, \dYdZ_l, D | V_\obs, \pi_\obs, \log \Teffobs, \FeH_\obs, \log g_\obs),  
\end{eqnarray}
where $\Delta M_i$ is the spacing between the $i$th and $i+1$th
stellar mass in our grid of stellar models, $\Delta \FeH_j$ is the
spacing between the $j$th and $j+1$th metallicity in our grid of
stellar models, $\Delta \dYdZ_l$ is the spacing between the $l$th and
$l+1$th value of $\delta Y/\delta Z$ in our grid of stellar models,
and $\Delta \tau _k$ is the spacing between the $k$th and $k+1$th age is
the spacing of the outputs in the stellar track with mass $M_i$ and
metallicity $\FeH_j$. The indicator function $I(R_o \le R \le R_o +
\Delta R)$ is defined to be 1 if $R_o \le R \le R_o + \Delta R$ and 0
otherwise.  For values of the various spacings between parameter
values in our grid of stellar models, see Table 1.  The stellar
evolutionary tracks are computed with a variable time step, so as to
efficiently evolve the star rapidly during the main sequence, but
provide the necessary temporal resolution to accurately model the
early and late stages of evolution.  In order to facilitate the
numerical integration, we interpolate within each stellar track to
provide each of the observable and derived physical quantities at a
series of stellar ages, each separated by $\Delta \tau _k = \Delta \tau$.
The value of $\Delta \tau$ varies between stars to reflect the speed of
stellar evolution for each track (see Table 1).  It is important to
note that we interpolate only in time within a single evolutionary
track, and not between stellar tracks of different masses or
compositions.  Thus, our interpolation does not suffer from any of the
complications typically associated with stellar isochrone fitting.
For a given set of observations, we can compute the marginal
quantities by summing the product of the prior times the likelihood
evaluated at each time of each evolutionary track.

In practice, the measurement uncertainties for each of the parameters
is much smaller than the entire allowed range of the parameter.
Therefore, we can approximate each of the above summations over all
model parameters by summations over a region $R'$ that contains all
the points that contribute significantly to the integrals.  We are
quite conservative, and choose $R'$ to include every point for which
$\left|\log \Teffobs - \log \Teffmod\right| < 10\sigma_{\Teff}$,
$\left|\log \FeH_\obs - \log \FeH_\mod\right| < 10\sigma_{\FeH}$,
$\left|\log g_\obs - \log g_\mod\right| < 10\sigma_{\log g}$, and
$\left|\pi_\obs -
10^{-\left(V_\obs-M_{V,\mathrm{mod}}(M,\tau,\FeH)\right)/5+1}\right| < 10
\sigma_\pi$.  We manually check that reducing the volume of $R'$
results in no significant difference for the marginal distributions.

\subsection{Characterizing the Derived PDFs \label{pdfanalysis}}

\begin{figure}
  \begin{center}
    \includegraphics[width=0.5\columnwidth]{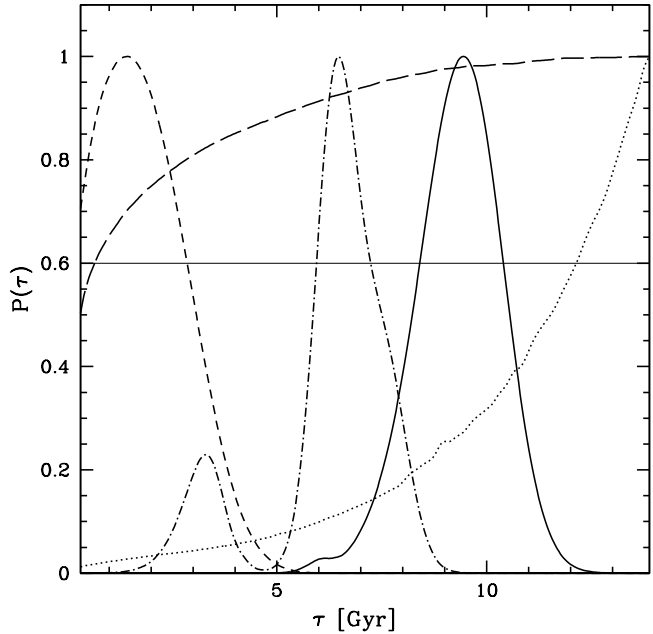}
		\caption{Normalized age probability distribution functions (PDFs) of sample stars.  HD101614 (solid curve) and HD193307 (dot-dashed curve) have ``well-defined'' ages.  HD193307 has a secondary maximum in the age-PDF, $\tau_{\rm 2}=3.4\,$Gyr with a relative probability of 23\,\% with respect to the best-estimate age, $\hat{\tau}=6.5\,$Gyr.  Only the best-estimate age and the upper-bound of the 68\,\% credible interval for the age of HD120237 (short-dashed curve) will be specified in the table ($\tau_{\rm high}=2.9\,$Gyr).  Only the lower-bound for the age of HD10700 (dotted curve) will be specified ($\tau_{\rm low}=12.2\,$Gyr).  No meaningful age can be derived for HD144628 (long-dashed curve).  The credible intervals are defined as the range between the points where the PDFs cross the solid horizontal line, $P(\tau)=0.6$ (see $\S\,$\ref{errorestimate}).  \label{samplePDF.fig}}
  \end{center}
\end{figure}

	While  we will provide the full posterior distributions upon request, it is often convenient to summarize the posterior PDFs.  Calculated PDFs are typically represented by two quantities --- a single ``{\it best estimate}'' value, and some associated credible interval.  However, the derived PDFs can often manifest complicated shapes that are not readily fit by a standard normal distribution (for example, see Figure~\ref{samplePDF.fig} for sample age-PDFs).  Thus, defining the best representative value and determining the credible interval is a non-trivial task.  An incorrect procedure for model parameter estimation can introduce an unwanted bias in a large sample and also fail to extract all the necessary information from the PDFs.  A useful representation of derived age probability distributions has been thoroughly discussed by \citet{jorgensen05}.
	
\subsubsection{Best Estimate \label{bestestimate}}

	The main goal of this section is to define a single estimate value $\hat{x}$, along with the range of plausible values $[x_1, x_2]$ determined from the selected credible level.  Commonly used statistical quantities for representing the best estimate from the PDFs are the median (the bisector of the area under the PDF curve), the mean (the expectation value), and the mode (the most probable value).  No matter which statistical quantity is selected, it inevitably includes certain arbitrariness and statistical biases.  Thus, when studying the statistical properties of a large sample of stars, one needs to be fully aware of the type of biases introduced and separate them carefully from the true values of interest.

	We have chosen the mode, i.e., the global maximum of the derived PDF, as our primary estimate of the model stellar parameter (denoted by the hat --- e.g., $\hat{\tau}$ for the best age-estimate).  The mean and the median have their advantages in that they always reside within any given parameter range.  For instance, the maximum age of HD10700 in Figure~\ref{samplePDF.fig} apparently lies beyond $14\,$Gyr, yet it is still possible to present the estimated age as $11.6\,$Gyr by choosing the median value.  However, as pointed out by \citet{jorgensen05}, when the derived PDF is broad due to a poor observation, the mean or the median tends to be deviously located near the middle of an arbitrary parameter range.  For example, in Figure~\ref{samplePDF.fig} it is in no sense meaningful to say that the estimated age of HD144628 to be $7.5\,$Gyr (the mean), or $7.7\,$Gyr (the median).  It is evident that these estimators would  lead to   a spurious bias toward $7\,$Gyr, the central value of our selected age interval, 0 -- 14\,Gyr,  when analyzing  a population of similar stars.  Furthermore, the calculated mass or age-PDFs often have multiple peaks as seen for HD193307 in Figure~\ref{samplePDF.fig}, since the model parameters ($M, \tau, Z, etc$) are not in one-to-one relation with the observed parameters ($\MV, \Teff, \FeH $).  Multiple observations for a single star can lead to bimodal posterior distributions that are even more difficult to summarize with a single estimator.  

\subsubsection{Error Estimate \label{errorestimate}}

	The conventional definition for $\pm 1\sigma$ credible interval is to assume a Gaussian distribution and find the range $[x_1,x_2]$ such that the fractional area under the curve falls to 68\% of the total integrated probability; $\int_{x_1}^{x_2} P(x) dx  /  \int_{x_{\rm min}}^{x_{\rm max}} P(x) dx = 0.68$.  Instead, we normalize the PDFs such that $P(\hat{x})=1.0$ and estimate our ``$1\sigma$ error'' $[x_1,x_2]$ to be the interval between the two points where $P(x_1) = P(x_2) = 0.6$ \citep{nordstrom04}.  This is a more appropriate choice for age-PDFs, since for many stars age is a particularly weakly  constrained parameter.  For example, the derived age-PDF for HD193307 in Figure~\ref{samplePDF.fig} has a smaller secondary maximum.  For such ambiguous cases, the credible interval estimated by the fractional area under the curve does not describe the correct uncertainty range for the best age estimate (the credible interval may even lie outside the mode value).  Also, computing the $1\sigma$ credible interval from 68\% fractional area  is likely to underestimate the uncertainty of broad age-PDFs (e.g., HD144628).  Note that for a standard Gaussian distribution, the credible interval $[x_1,x_2]$ defined as $P(x_1) = P(x_2) = 0.6$ is the region in which the true parameter lies for 68\% of all the cases.  This definition of a credible interval can be also applied to non-Gaussian type model parameter PDFs.  \citet{jorgensen05} have done extensive Monte Carlo simulations using $10^3$ synthesized stars with typical observational errors, and confirmed that 68\% of the recovered ages fell within $[x_1,x_2]$ of the true age.

\subsubsection{Our Notation for the Parameter Estimates and the Credible Intervals \label{notation}}

	Here we summarize the notations we have used to characterize the parameter estimates and credible intervals in Table~\ref{modelparameters.tab}.  When the credible interval around the mode is entirely within the parameter range of our grids, $[x_{\rm min}, x_{\rm max}]$,  the estimate is called a ``well-defined'' parameter \citep{nordstrom04}.  However, it often happens, particularly in the age-PDFs, that the credible interval is truncated on one or both sides of $[x_{\rm min}, x_{\rm max}]$ (e.g., HD120237, HD10700 and HD144628 in Figure~\ref{samplePDF.fig}).  In these cases, the truncated side of the credible interval is left blank in the table.  This means that only an upper or lower bound (or, possibly, no bound at all) can be specified at the 68\% credible level for the considered parameter.  Similarly, sometimes the mode is not well-defined, as the position of the global maximum $\hat{x}$ coincides with either boundary of the permitted parameter range  (e.g., $P(x_{\rm max})$ is unity for HD144628).  In this case, the best-estimate value $\hat{x}$ is not specified in the table.  However, an upper or lower bound of the credible interval may still be specified, if it exists.  

	Lastly, the calculated model parameter PDFs often exhibit a bimodal feature (e.g., HD193307).  Sometimes the secondary maximum lies outside the credible interval of the primary.  In these cases, the secondary maximum and its probability relative to the global maximum are also specified in the table.  

	It is often useful to also provide summary statistics for each of the model parameters.  By their very nature, summary statistics throw away much of the information that is contained in the posterior PDFs.  The loss of information is particularly acute for complex PDFs (e.g., very broad, highly skewed, multi-modal), such as often occur with the marginal posterior PDF for the stellar age. In order to provide a succinct summary of the size of the tails of the posterior PDFs, we provide the median value as well as 68\,\%, 95\,\% and 99\,\% credible intervals for each stellar parameter in the electronic version of the table.
\section{Derived Stellar Properties \label{results}}  
	
	With the methods described in the previous chapter, we have computed the stellar parameter PDFs for 1074 SPOCS stars with $T_{\rm eff}$ within the range 3700 -- 6900\,K.  The derived stellar parameters for a small subset of the sample are presented in Table~\ref{modelparameters.tab}.  The entire table of derived parameters is available in the electronic version of the paper and also in the California \& Carnegie Planet Search website \footnote{http://exoplanets.org/SPOCS\_evol.html}.  

\input{tab2}
\begin{figure*}[htb]
  \begin{center}
	\plottwo{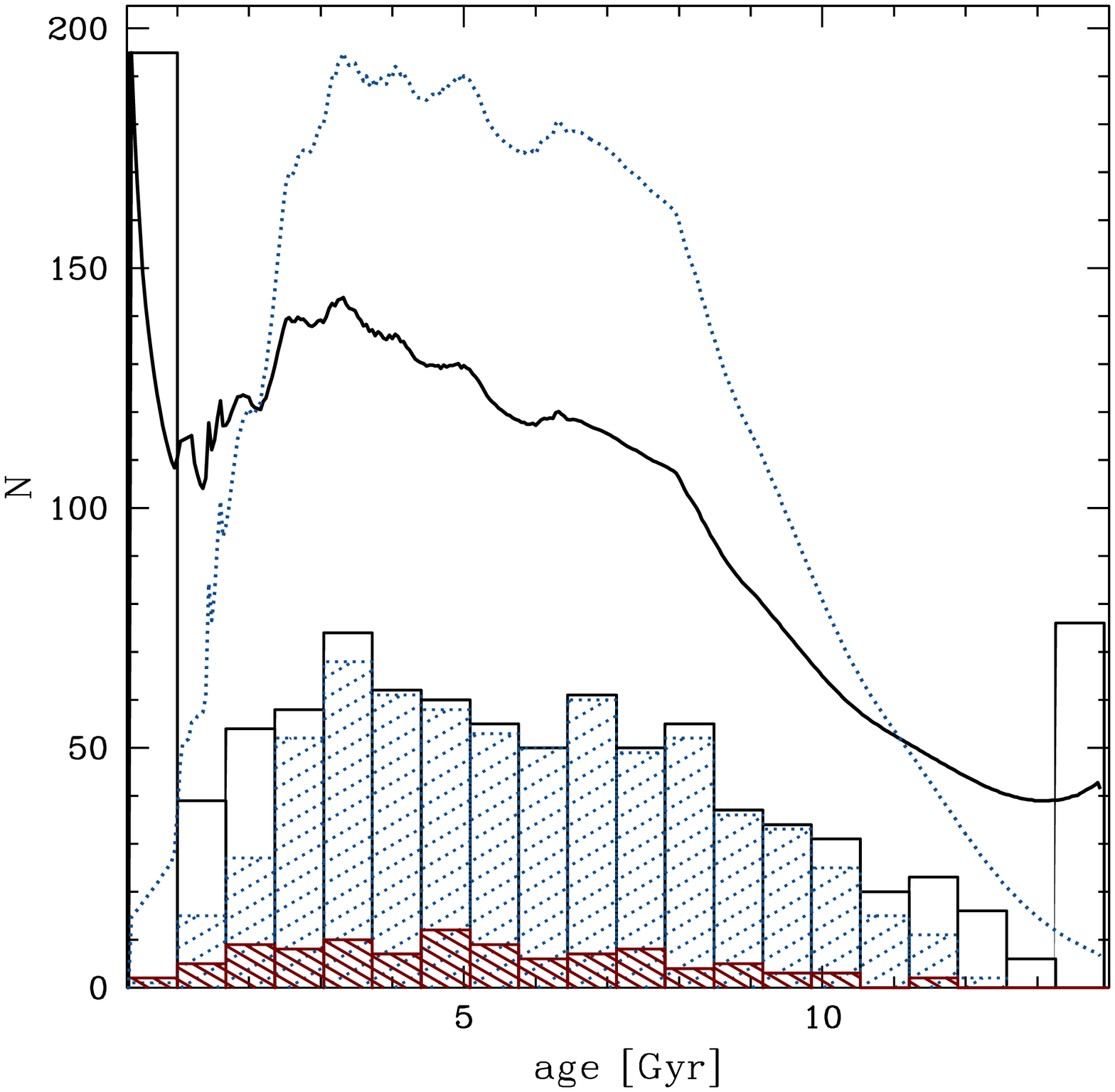}{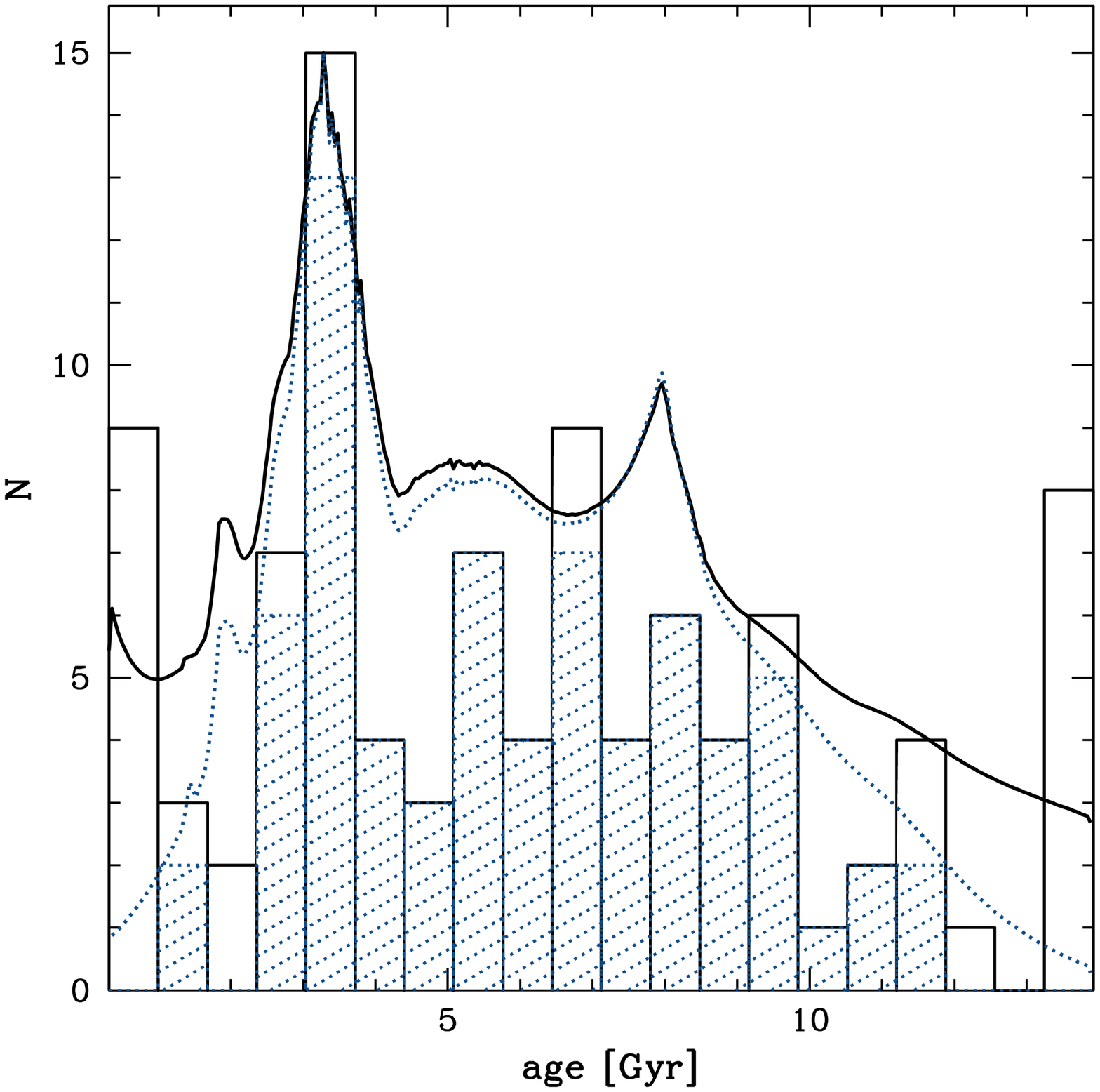}
		\caption{Distributions of derived ages.  The right panel includes only the 99 stars in the SPOCS catalog with known planetary companions.  The best-estimate ages are presented in the histograms, and the integrated total age probability distribution functions are presented as the curves.  The blue dotted line includes only the stars with well-defined ages.  The red dashed curve is the age distribution of the 61 stars in the volume-limited sample with well-defined ages.  \label{age.fig}}
  \end{center}
\end{figure*}
\begin{figure}
  \begin{center}
    \includegraphics[width=0.5\columnwidth]{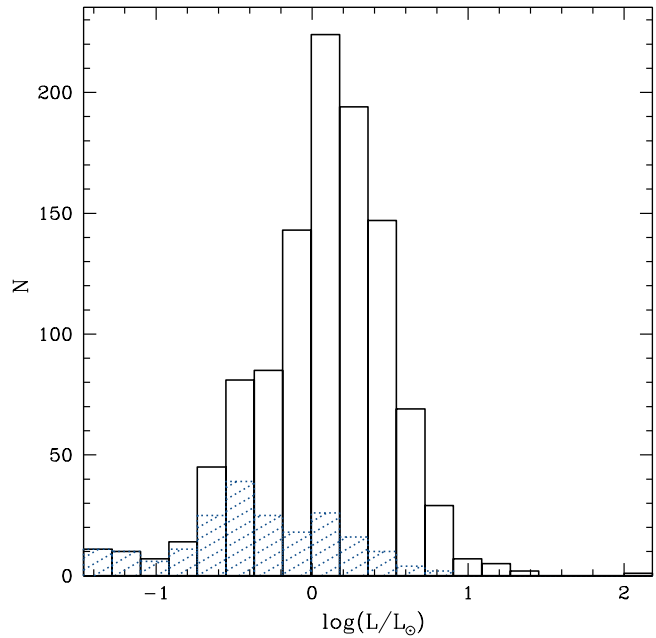}
		\caption{Distributions of the calculated luminosity for the whole sample (solid) and for the 203 stars in the volume-limited sample (dotted, blue).  The volume-limited sample has a luminosity distribution peaked around $\log{L/L_\odot}=-0.32$ whereas the luminosity of all the sample stars are peaked at $\log{L/L_\odot}=0.15$.  A larger sampling volume includes more intrinsically bright stars. \label{L.fig}}
  \end{center}
\end{figure}

\subsection{Stellar Ages}
	Figure~\ref{age.fig} shows the derived age distributions of stars in different sub-samples of the SPOCS catalog.  The curves show the sum of the normalized age-PDFs of all the stars.  Among the total of 1074 SPOCS stars we have used, 669 stars have ``well-defined'' ages as defined in $\S\,$\ref{notation}.  Among those, 606 stars have estimated errors smaller than $5\,$Gyr, and 169 stars have errors smaller than $1\,$Gyr.  The large fraction of stars with the youngest ages ($\tau<1\,$Gyr) and the oldest ages ($\tau>13\,$Gyr) in Figure~\ref{age.fig} are clearly artifacts from choosing the mode value as the best-estimate age.  As discussed in $\S\,$\ref{bestestimate}, the mode of a poorly constrained age-PDF tends to reside near either end of the selected age range (0 or $14\,$Gyr in this case).  These accumulations at extreme ages are removed in the distribution of ``well-defined'' ages.  

	Note that the selection criterion of the SPOCS catalog is such that precision of  radial-velocity observations is optimized.  The sample stars are selected by visual magnitude, but not the stellar distance.   Consequently, the catalog contains more distant, early-type stars that are intrinsically more luminous.  \citet{fischer05} defined a volume-limited sample with a radius of $18\,$pc, within which the number density of F-, G- and K-type stars per volume is nearly constant as a function of distance.  Among the 1074 stars we selected from the SPOCS catalog, 203 stars are in the volume-limited sample.  Figure~\ref{L.fig} illustrates the bias toward intrinsically luminous stars in the whole sample.  The volume-limited sample is in fact rather abundant in intrinsically faint stars, whereas the whole sample consists of more stars with luminosity above the solar value.  The age distribution of the volume-limited sample is slightly shifted toward a younger age, due to a larger fraction of relatively unevolved, less bright stars.  
	
	The age distribution of stars with known planetary companions is presented in the right panel of Figure~\ref{age.fig}.  The SPOCS catalog contains 99 planet-host stars, 74 of which have well-defined ages.  The integrated age-PDF of all the stars has sharp local maxima, arising from the small number statistics and selection effects.  The mode value of $3.3\,$Gyr of the planet-host stars coincides with that of all the well-defined ages, however, the peak is more distinct for the planet-host stars.  The two maxima around $3.3\,$Gyr and $8.0\,$Gyr mostly consist of G-type stars, since planet-search programs typically target stars with solar-type spectra.  Most of the stars in the $2-5\,$Gyr bins consist of young main-sequence stars with spectral type F8 -- G4 and masses 1.0 -- $1.9\,M_\odot$, whereas the stars in the 7 -- 9\,Gyr bins are less massive G3 -- G5 stars with masses 0.8 -- $1.1M_\odot$, many of them likely to be in the subgiant phase.   

	Note that the stellar age is generally the most poorly-determined parameter.  Nearly half of the SPOCS stars have derived age uncertainty greater than 5\,Gyr, which is comparable to the entire parameter range (0 -- 14\,Gyr).  Also, the age uncertainty is more sensitive to the stellar mass than on the accuracy of the spectroscopic observations (the derived age -- mass relation is discussed in more details in \S~\ref{age-M}).  Figure~\ref{age.fig} merely provides a graphical representation of the summary statistics and the contributions from different types of stars to the overall age distribution, and it should not be confused with the actual star-formation rate in the solar neighborhood.

\begin{figure*}
  \begin{center}
	\plottwo{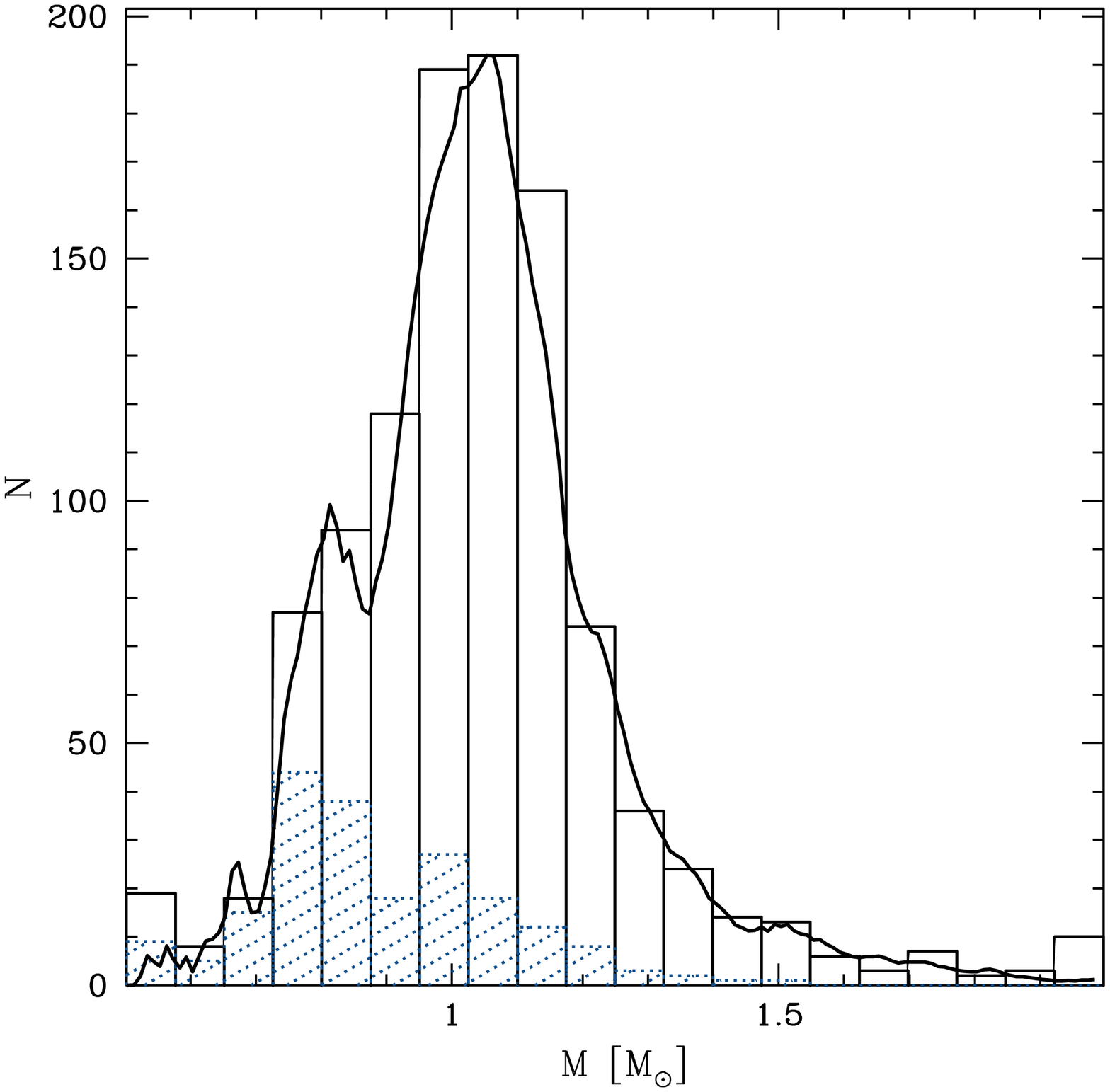}{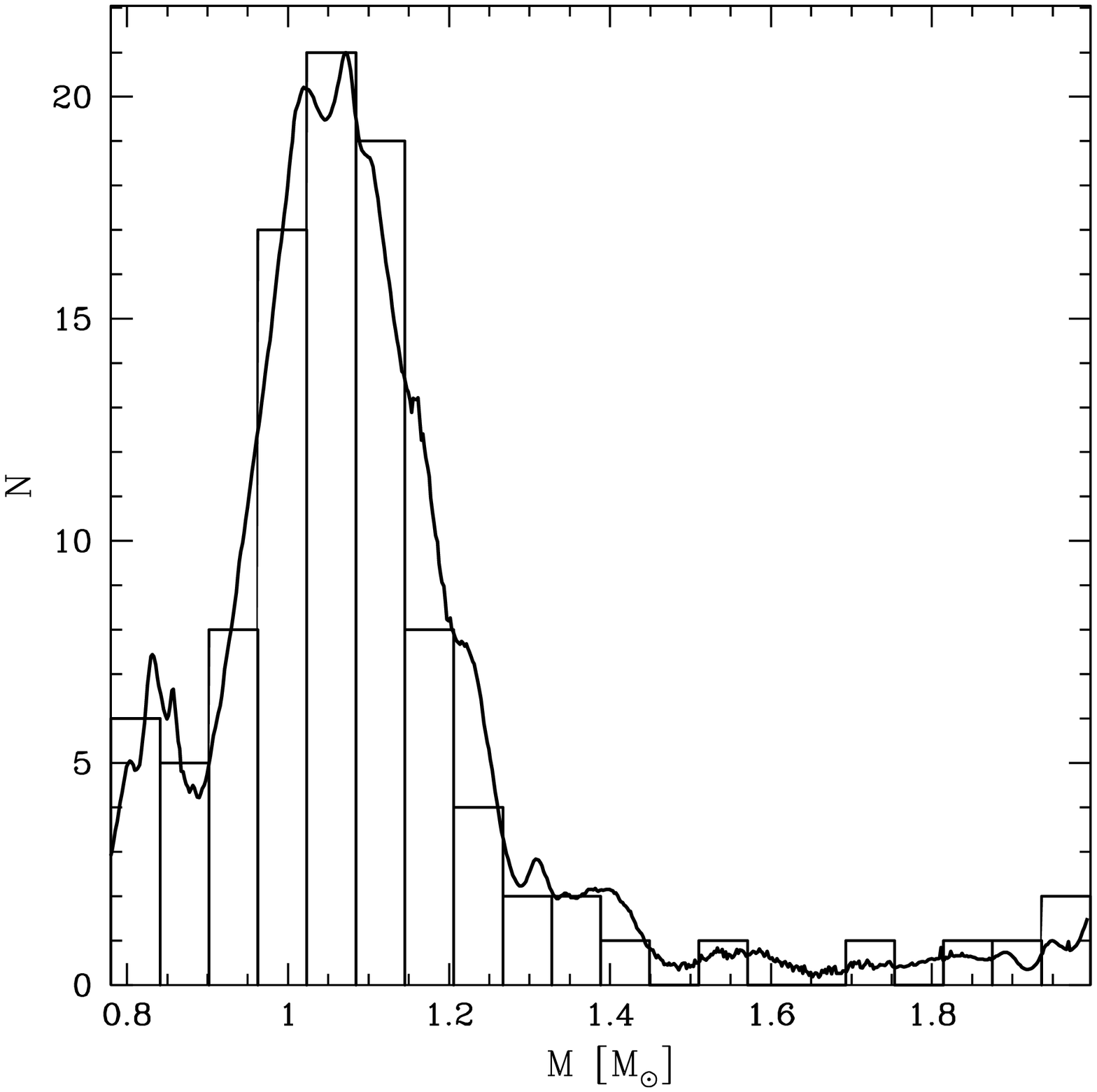}
		\caption{Distributions of the derived masses.  The right panel includes only the 99 stars in the SPOCS catalog with known planetary companions.  The best-estimate masses are presented in the histograms, and the integrated total mass probability distribution functions are presented as the curves.  The mass distribution for the stars in the volume-limited sample ($<18\,$pc) is presented as the blue dotted curves .  \label{M.fig}}
  \end{center}
\end{figure*}
\begin{figure*}
  \begin{center}
    \includegraphics[width=0.5\columnwidth]{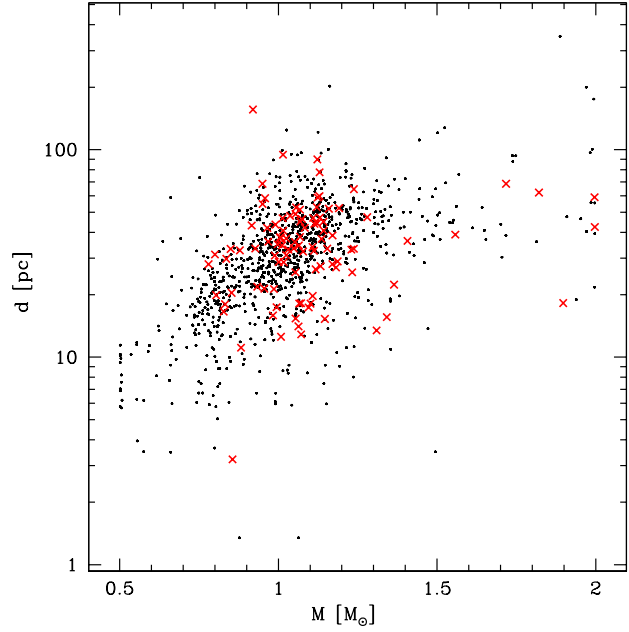}
		\caption{Distribution of the derived masses and the distances determined from the Hipparcos parallaxes.  The red crosses represent the stars with known planetary companions.  \label{M_d.fig}}
  \end{center}
\end{figure*}

\subsection{Stellar Masses}
	Figure~\ref{M.fig} shows the distributions of derived masses.  The mass-PDFs are generally much better constrained than the age-PDFs.  Well-defined masses are obtained for the majority (97\%) of sample stars, including the ones with poorly determined ages.  The derived masses have a median of $1.03M_\odot$, consistent with the median of $1.07M_\odot$ quoted by VF05 within the $1\sigma$ uncertainty.  Since only the stars with larger masses and luminosities are selected at large distances, there is a clear correlation between the derived mass and the stellar distance (see Figure~\ref{M_d.fig}).  The volume-limited sample contains more stars with sub-solar masses, with a median value of $0.87M_\odot$.  
	
	The small gap in the distribution around $0.9M_\odot$ is likely to originate from our choice of age cutoff at $14\,$Gyr.  The mass of $0.9M_\odot$ roughly corresponds the turn-off mass for this age (the critical mass beyond which a star can evolve away from the main sequence within $14\,$Gyr; see, for example, the sample evolutionary tracks for 55~Cancri in Figure~\ref{55Cnc.fig}).  Indeed, the 125 stars in the mass range 0.8 -- $0.9M_\odot$ have particularly old ages: 39 stars (31\%) with ages $>12\,$Gyr and 58 stars (46\%) with $>10\,$Gyr.  Note that in the complete sample only 7\% of the stars have derived ages $\tau >12\,$Gyr, and 16\% have  $\tau >10\,$Gyr.
	
	The mass distribution of the 99 known planet-host stars is also presented in the right panel of Figure~\ref{M.fig}.  The mass distribution of planet-host stars behaves similarly to that of the entire sample, with a median mass of $1.07M_\odot$.  The lower-mass cutoff near $\sim0.8M_\odot$ comes from the sampling criterion ($V<8.5$) of the SPOCS catalog.  Also, stars with masses below $\sim0.8M_\odot$ correspond to M-dwarfs, which are typically excluded from planet-search programs because of their low luminosities and strong chromospheric activities, which affect the radial-velocity measurements \citep{delfosse00}.  All the planet-host stars with derived masses $< 0.9M_\odot$ in our sample are K0 -- 3 stars.  There are only four stars in the extrasolar planet catalog with spectra later than K3: HD63454 (K4V), GJ436 (M2.5), Gl581 (M3) and GJ876 (M4V).  None of these stars is included in the SPOCS catalog.    
	
\begin{figure*}
  \begin{center}
    \includegraphics[width=\columnwidth]{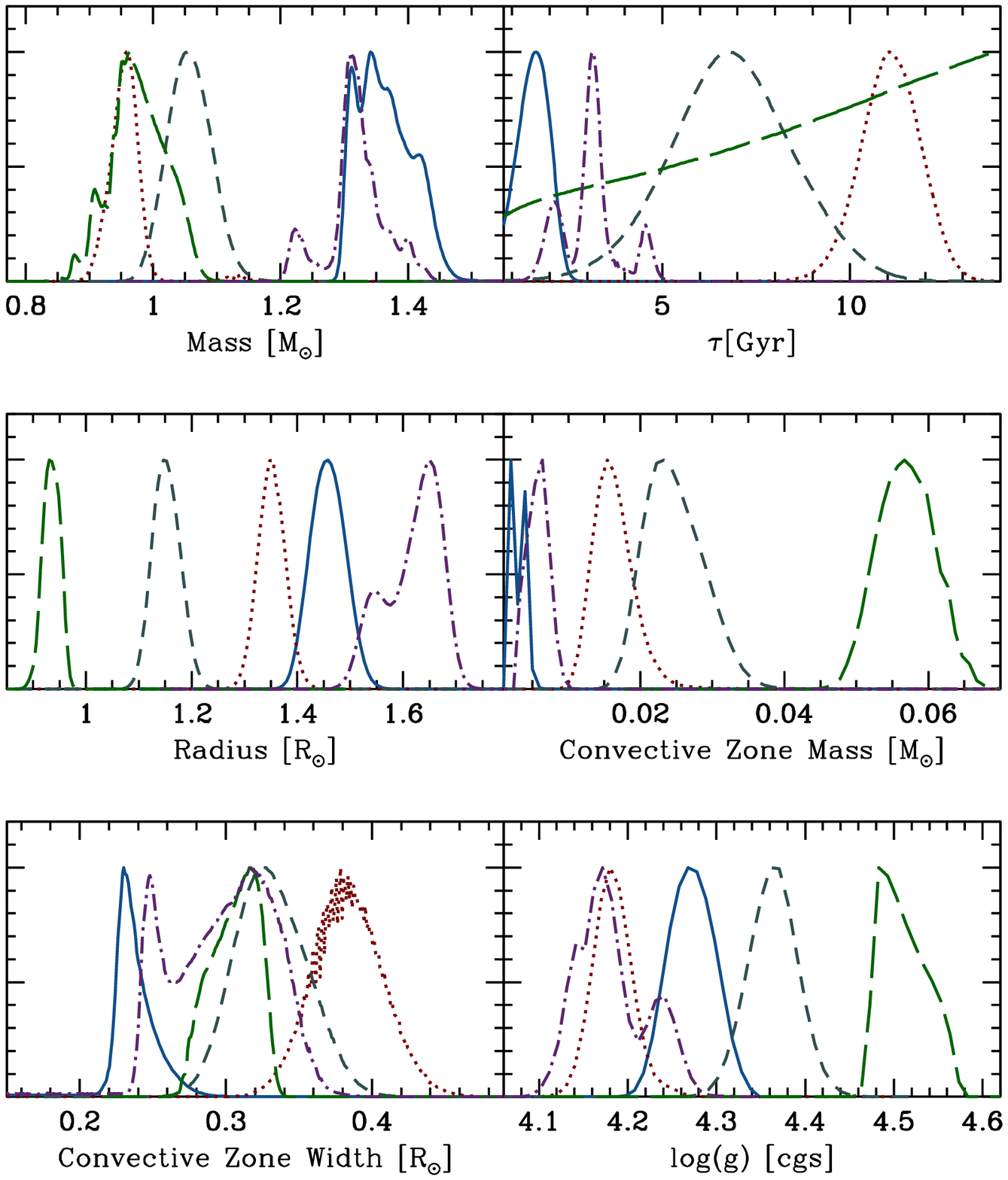}
		\caption{The derived PDFs for mass ($M$), age ($\tau$), radius ($R$), mass and width of the convective zone ($M_{\rm ce}, R_{\rm ce}$) and surface gravity ($\log{g}$) of five planet-host stars: $\tau\,$Boo (solid, blue), 51~Peg (dark-green, short-dashed), $\upsilon\,$And (purple, dot-dashed), 55~Cnc (green, long-dashed) and $\rho\,$CrB (red, dotted).  The derived parameters and credible intervals are listed in Table~\ref{comparison_ford.tab}\label{five.fig}. }
  \end{center}
\end{figure*}
\begin{deluxetable*}{lrrrrrrr}
\tablecolumns{8}
\tablewidth{0pc}
\tablecaption{Properties of Sample Stars \label{obsandmodel.tab}}
\tablehead{
\colhead{}	&	\multicolumn{3}{c}{Observed Data}	&	\colhead{}	&	\multicolumn{3}{c}{Posterior} \\
\cline{2-4} \cline{6-8} \\
\colhead{Star}	&	\colhead{$T_{\rm eff}$ [K]}	&	\colhead{$[{\rm Fe/H}]$}	&	\colhead{$\log{g}$ [cgs]}	&	\colhead{}	&	\colhead{$T_{\rm eff}$ [K]}	&	\colhead{$[{\rm Fe/H}]$}	&	\colhead{$\log{g}$ [cgs]}}
\startdata
$\tau$ Boo & 6387$\pm$44 & +0.25$\pm$0.03 & 4.25$\pm$0.06 &
& 6390$^{-73}_{+74}$ & +0.31$\pm$0.04 & 4.27$^{-0.02}_{+0.04}$ \\
51 Peg & 5787$\pm$44 & +0.15$\pm$0.03 & 4.45$\pm$0.06 &
& 5814$^{-53}_{+67}$ & +0.22$\pm$0.03 & 4.36$^{-0.02}_{+0.04}$ \\
$\upsilon$ And & 6213$\pm$44 & +0.12$\pm$0.03 & 4.25$\pm$0.06 &
& 6159$^{-42}_{+71}$ & +0.16$\pm$0.04 & 4.17$^{-0.03}_{+0.02}$ \\
55 Cnc & 5253$\pm$44 & +0.31$\pm$0.03 & 4.45$\pm$0.06 &
& 5327$\pm49$ & +0.37$\pm$0.04 & 4.48$^{-0.01}_{+0.05}$ \\
$\rho$ CrB & 5823$\pm$44 & -0.14$\pm$0.03 & 4.36$\pm$0.06 &
& 5855$^{-54}_{+81}$ & -0.17$\pm$0.05 & 4.21$^{-0.02}_{+0.05}$ \\
\enddata
\tablecomments{The surface properties of the sample stars  --- comparison between the spectroscopic values by VF05 and the posterior estimates. } 
\end{deluxetable*}
\begin{deluxetable*}{lrrrrrrr}
\tabletypesize{\scriptsize}
\tablecaption{Comparisons with Ford et al. \label{comparison_ford.tab}}
\tablewidth{0pt}
\tablehead{
	\colhead{Star} & \colhead{$M_* [{\rm M}_\odot]$} & \colhead{Age [Gyr]} & \colhead{$R_* [{\rm R}_\odot]$} &\colhead{$M_{\rm ce} [{\rm M}_\odot]$} &\colhead{$R_{\rm ce} [{\rm R}_\odot]^a$} & \colhead{$\log{g}$ [cgs]} 
}
\startdata
$\tau$ Boo (HD120136) & 1.34$^{-0.04}_{+0.05}$ & 1.64$^{-0.52}_{+0.44}$ & 1.46$\pm$0.05 & 0.002$^{-0.002}_{+0.003}$ & 0.23$\pm$0.01 & 4.27$^{-0.03}_{+0.04}$ \\
 & 1.37$\pm0.08$ & 1.2$^{-0.8}_{+1.2}$ & 1.41$^{-0.09}_{+0.10}$ & $\la0.002$ & 0.22$^{-0.18}_{+0.19}$ & 4.27$^{-0.07}_{+0.05}$ \\

51 Peg (HD217014) & 1.05$\pm$0.04 & 6.76$^{-1.48}_{+1.64}$ & 1.15$\pm0.04$ & 0.023$^{-0.004}_{+0.006}$ & 0.33$^{-0.02}_{+0.03}$ & 4.36$^{-0.03}_{+0.04}$ \\
 & 1.05$^{-0.08}_{+0.09}$ & 7.6$^{-5.1}_{+4.0}$ & 1.16$\pm0.07$ & 0.023$^{-0.006}_{+0.007}$ & 0.34$\pm0.11$ & 4.33$\pm0.09$ \\

$\upsilon$ And (HD9826) & 1.31$^{-0.01}_{+0.02}$ & 3.12$^{-0.24}_{+0.20}$ & 1.64$^{-0.05}_{+0.04}$ & 0.005$^{-0.003}_{+0.002}$ & 0.32$^{-0.07}_{+0.03}$ & 4.16$^{-0.04}_{+0.02}$ \\
 & 1.34$^{-0.12}_{+0.07}$ & 2.6$^{-1.0}_{+2.1}$ & 1.56$^{-0.10}_{+0.11}$ & 0.002$^{-0.002}_{+0.003}$ & 0.27$^{-0.10}_{+0.17}$ & 4.18$^{-0.10}_{+0.07}$ \\

55 Cnc (HD75732) & 0.96$^{-0.03}_{+0.05}$ & $>7.24$ & 0.93$^{-0.01}_{+0.03}$ & 0.057$^{-0.004}_{+0.005}$ & 0.32$^{-0.03}_{+0.01}$ & 4.48$^{-0.01}_{+0.05}$ \\
 & 0.95$^{-0.09}_{+0.11}$ & 8.4$^{-8.3}_{+7.1}$ & 0.93$^{-0.03}_{+0.02}$ & 0.046$^{-0.004}_{+0.006}$ & 0.30$^{-0.06}_{+0.05}$ & 4.50$^{-0.07}_{+0.04}$ \\

$\rho$ CrB (HD143761) & 0.96$\pm0.02$ & 11.04$^{-0.72}_{+0.88}$ & 1.35$^{-0.02}_{+0.03}$ & 0.015$^{-0.002}_{+0.003}$ & 0.38$\pm0.02$ & 4.18$^{-0.01}_{+0.03}$ \\
 & 0.89$^{-0.04}_{+0.05}$ & 14.1$^{-2.4}_{+2.0}$ & 1.35$^{-0.08}_{+0.09}$ & 0.033$^{-0.009}_{+0.011}$ & 0.47$^{-0.13}_{+0.12}$ & 4.13$^{-0.06}_{+0.07}$ \\

\enddata
\tablecomments{Comparisons with the calculations from Ford et al. (1999).  Their results are shown in the second row of each star.}
\tablenotetext{a}{Width of the convective zone, measured from the outermost stellar surface}

\end{deluxetable*}
	

\section{Discussion}
	
\subsection{Comparisons with Paper I. \label{comparison}}
		
 FRS99 modeled five stars with extrasolar planets known at the time, using the observed parameters obtained by \citet{gonzalez97,gonzalez98a} and \citet{gonzalez98b}.  In this section, we compare those results to our new models using the spectroscopic observations by VF05.  The overall comparisons are summarized in Table~\ref{comparison_ford.tab}, and the theoretically derived atmospheric properties are listed in Table~\ref{obsandmodel.tab}.  The normalized PDFs of the stellar properties are presented in Figure~\ref{five.fig}.

\subsubsection{$\tau$ Bootis (HD120136)}
A planet with a mass $\Mpl \sin{i}\approx3.9M_{\rm J}$ around the young F7V star $\tau\,$Boo was discovered by \citet{butler97} in a 3.31-day period.  Despite its assumed young age, the star shows little photometric variability \citep{baliunas97}.  Among the five planets discussed in this section, it is the hottest and the second most metal-rich star after 55~Cnc.  The mean effective temperature $T_{\rm eff} = 6410\,$K and metallicity $[{\rm Fe/H}] = +0.26\,$ are derived by VF05 using the SME with Lick/Hamilton spectra at three different epochs.  The planetary system $\tau\,$Boo is associated with a visual binary companion at $\sim240\,$AU \citep{hale94,patience02}.  The stellar companion does not affect the spectral analysis because of  its large orbital separation.

Our calculated properties of $\tau\,$Boo are consistent with the model made by FRS99:  $\tau\,$Boo is a young, fairly massive main-sequence star with a well-constrained age estimate, $1.64^{-0.52}_{+0.44}\,$Gyr.  The activity -- age relation determined from the mean Ca II flux suggests an age of $\sim2\,$Gyr \citep{baliunas97}.  The revised spectroscopic analysis by \citet{gonzalez00} has been applied to isochrones by \citet{schaller92} and \citet{schaerer93} and yielded an age $3.3\pm0.5\,$Gyr.  All the other isochrone or surface activity analyses estimate the age to be $<3.1\,$Gyr, consistent with our results \citep{fuhrmann98,lachaume99,suchkov01,henry00b,nordstrom04}.  

\begin{figure}
  \begin{center}
    \includegraphics[width=0.4\columnwidth]{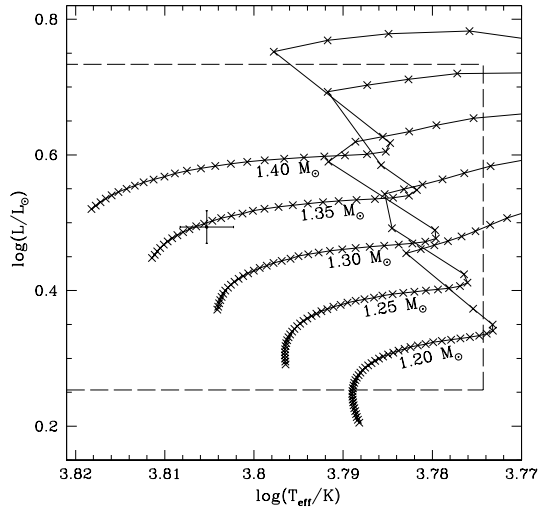}
		\caption{Theoretical HR diagram and stellar evolutionary tracks for $\tau\,$Bootis.  The observed $T_{\rm eff}$ from the SPOCS catalog and luminosity calculated from the observed magnitude and Hipparcos parallax are shown with $1\sigma$ uncertainty.  All the evolutionary tracks inside the dashed rectangle ($10\sigma$ observational uncertainty) are used for the stellar model calculation.  The tracks are presented from ZAMS onward, with our best-fit metallicity $\FeH=+0.32$.   Each $\times$ on the tracks is separated by $100\,$Myr.  Note that this time step is exaggerated in this figure to illustrate the acceleration of the evolutionary sequences --- in the actual grids used for the calculations, the time resolution is as fine as $1\,$Myr.  Theoretically derived model parameters of $\tau\,$Boo are $M=1.34M_\odot$ and age $\tau = 1.64\,$Gyr.  \label{tauBoo.fig}}
  
  \end{center}
\end{figure}

Figure~\ref{tauBoo.fig} shows the theoretical evolutionary tracks with $[{\rm Fe/H}] = +0.31\,$.  This metallicity estimate is consistent with other spectroscopic observations: $+0.32\pm0.06\,$ \citep{gonzalez00} and $+0.27\pm0.08\,$ \citep{fuhrmann98}.  VF05 and \citet{santos04} have derived slightly lower values, $0.25\pm0.03\,$ and $0.23\pm0.07\,$, respectively.

\subsubsection{51 Pegasi (HD217014)}

The first extrasolar planet around a solar-type star was discovered by \citet{mayor95}, around the G5V star 51 Pegasi.  The planet has a mass $\Mpl \sin{i}\approx0.47\MJ$ and an orbital period of 4.23 days.  

From the three Lick/Hamilton spectra, VF05 derived $T_{\rm eff}=5787\pm44\,$K and $[{\rm Fe/H}]=0.154\pm0.029\,$, consistent with other LTE analyses, $T_{\rm eff}=5750$ -- $5820\,$K and $[{\rm Fe/H}]=0.14$ -- $0.21\,$ \citep{gonzalez98b,gimenez00,henry00b,santos04,chen06,ecuvillon06}.  Our theoretical model of 51~Peg implies that it may be slightly hotter and more metal-rich than the best-fit parameters obtained by VF05 (see Table~\ref{obsandmodel.tab}).  The best metallicity estimate for 51~Peg calculated from the theoretical tracks is $[{\rm Fe/H}]=+0.22\,$.  The star is slightly older than the Sun, $6.76^{-1.48}_{+1.64}\,$Gyr old.  This is in good agreement with FRS99 ($7.6^{-5.1}_{+4.0}\,$Gyr) and most of other isochrone analyses: $4\pm2\,$Gyr \citep{fuhrmann98}, $9.2^{-4.4}_{+2.8}\,$Gyr \citep{nordstrom04} and $5.1^{-0.8}_{+3.0}\,$Gyr \citep{lachaume99}.  From the re-calibrated stellar rotation period determined from the Ca~II observation, \citet{henry00b} yielded an age of 3 -- 7\,Gyr.  The derived mass, age, radius and the mean rotational period \citep[$\sim25\,$days,][]{henry00b} all indicate that 51~Peg is a star very similar to our Sun, except for its metal-rich atmosphere.

\begin{figure}
  \begin{center}
    \includegraphics[width=0.4\columnwidth]{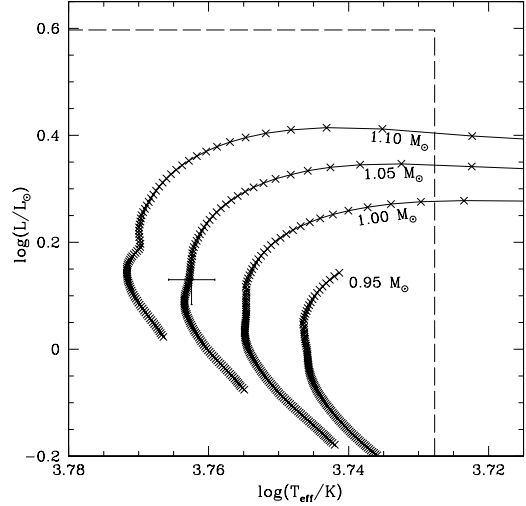}
		\caption{Theoretical HR diagram and stellar evolutionary tracks for 51~Pegasi.  The tracks correspond to our best estimate metallicity, $[{\rm Fe/H}] = +0.22\,$.  The track for $0.95M_\odot$ is truncated at $14\,$Gyr.  The stellar mass and age estimates are $1.05M_\odot$ and $6.76\,$Gyr.  The star is similar to our Sun except for its metal-rich atmosphere.  \label{51Peg.fig}}
  \end{center}
\end{figure}

\subsubsection{$\upsilon\,$ Andromedae (HD9826)}
\begin{figure}
  \begin{center}
    \includegraphics[width=0.4\columnwidth]{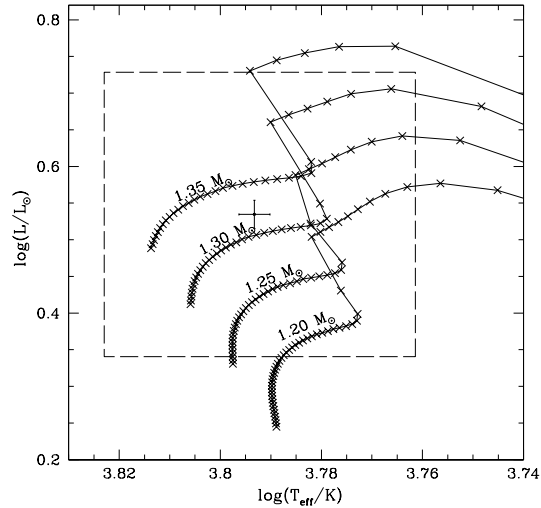}
		\caption{A small subset of theoretical evolutionary tracks used for the modeling of $\upsilon\,$Andromedae.  The temperature and luminosity quoted by VF05 are denoted with $1\sigma$ error bar and $10\sigma$ error box.  The tracks represent the evolutionary sequences of stars with metallicity $\FeH=+0.20$, corresponding to our best-estimate metallicity.  The star is most likely a $1.31M_\odot$ main-sequence star that is 3.12\,Gyr old.  Another model with $M\la1.25M_\odot$ at a main-sequence turn-off age ($\sim5\,$Gyr) is also possible, although with a much smaller probability.  \label{upsAnd.fig}}
  \end{center}
\end{figure}

The bright F8V star, $\upsilon\,$And is a known triple-planet system, harboring three Jupiter-size planets at $0.059\,$AU, $0.829\,$AU and $2.53\,$AU \citep{butler97,butler99}.  $\upsilon\,$And has a distant sub-stellar companion (M4.5V) at $\sim750\,$AU revealed by co-proper motion \citep{lowrance02}, which does not affect  spectral analysis.  Four Lick/Hamilton spectra were obtained for $\upsilon\,$And by VF05.  The atmospheric properties derived from each spectrum show a modest range of temperature and metallicity, $\Teff = 6150$ -- 6334\,K and $\FeH=0.08$ -- 0.192.  In general, modeling higher-mass stars involve some ambiguity as the sharp hook starts to appear at the end of the main sequence following the core hydrogen exhaustion (cf. $\S\,$\ref{grids} and Figure~\ref{upsAnd.fig}).  The discrepancies in the spectroscopically determined atmospheric properties is sensitively reflected in the stellar models, as seen in the derived PDFs. 

Overall, our theoretically derived properties agree with those by FRS99.  The derived age-PDF indicates that $\upsilon\,$And is $3.12^{-0.24}_{+0.20}\,$Gyr old, with two additional probability peaks around it.  The majority of the spectral and photometric analyses agree with the young age: using theoretical isochrones, $2.9\pm0.6\,$Gyr \citep{lachaume99}, $3.3^{-0.7}_{+1.7}\,$Gyr \citep{nordstrom04}, $3.8\pm1.0\,$Gyr \citep{fuhrmann98}; using age -- activity relation from Ca II flux observation, $5\,$Gyr \citep{donahue93}; using the age -- [Fe/H] relation, an upper bound of $2.3\,$Gyr \citep{saffe05}.  Another possible model with a relative probability 20 -- 40\,\% is observed in the calculated PDFs, which has a higher mass ($\sim1.4M_\odot$) and even younger age ($\sim2\,$Gyr).  This is attributed to one of the Hamilton spectra that yields hotter surface temperature and higher metallicity than the other three ($6334\,$K and $+0.192\,$).  The SPOCS catalog denotes the mean values of all four observations, $T_{\rm eff}=6213\,$K and $[{\rm Fe/H}]=+0.122\,$, consistent with the other reported values, $6107-6212\,$K and $0.09$ -- 0.17 \citep{fuhrmann98,gimenez00,gonzalez00,santos04}.  With a lower probability, $\upsilon\,$And could also be a near turn-off star with a lower mass ($\sim1.25M_\odot$) and an older age ($\sim4.5\,$Gyr), not because of the observational ambiguity but because of the sharp rise in the stellar luminosity after core hydrogen exhaustion.  This probability is relatively smaller due to the rapid evolution of the star away from the main sequence toward the lower temperature.  As mentioned above, all other isochrone analyses with appropriate statistical treatment (e.g., Bayesian analysis) favor the main-sequence model younger than $3.5\,$Gyr, rather than turn-off or post-main-sequence models.  

\subsubsection{55 Cancri (HD75732)}

55\,Cnc is the richest planet-host star known to date, harboring four Neptune- to Jupiter-size planets \citep{butler97,marcy02,mcarthur04}.  This is a system of great interests for many aspects of planetary dynamics.  The innermost planet 55\,Cnc\,e is a hot sub-Neptune mass planet, while the outermost planet 55\,Cnc\,d is one of the few planets known with an orbital semimajor axis comparable to that of Jupiter ($a=5.257\,$AU).  The middle two planets are in 3:1 mean motion resonance \citep{ji03,marzari05}, posing an interesting question for dynamical stability.  55\,Cnc is also a visual binary system, with a stellar companion at $\sim1100\,$AU away from the planet-host star \citep{hoffleit82,mugrauer06}.  In addition, the star was once claimed to have a Vega-like dust disk based on infrared observations \citep{dominik98,trilling98}.  Observations with different wavelengths have set an upper limit on the disk mass much lower than the previous estimate from the infrared observation, and now it is generally agreed that the infrared excess
most likely came from a background object \citep{jayawardhana02,schneider01}.

\begin{figure}
  \begin{center}
    \includegraphics[width=0.4\columnwidth]{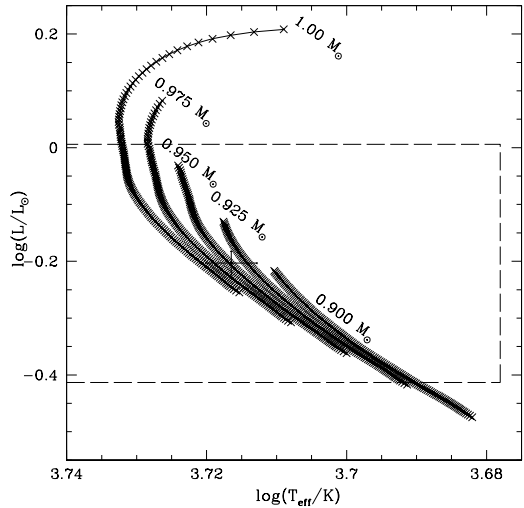}
		\caption{Observed temperature and luminosity of 55~Cancri and the evolutionary tracks of stars with the best-estimate metallicity $\FeH=+0.38$.  All the tracks are terminated at $14\,$Gyr.  Effective temperatures determined from other spectroscopic observations typically lie within a range,    $\log{(\Teff / K)}=3.70$ -- 3.73.  The star is most likely a main-sequence star with a mass $0.96^{-0.03}_{+0.05}M_\odot$  \label{55Cnc.fig}}
  \end{center}
\end{figure}

The strong photospheric absorption lines of 55\,Cnc indicate an anomalous metal abundance, classifying the star as a so-called ``super-metal-rich'' star.  The peculiar spectrum of 55\,Cnc results in controversial atmospheric properties.   The star is normally classified as a G8V star \citep{cowley67}, whereas \citet{taylor70} identified it as a super-metal-rich K~dwarf.  Reported surface temperature and metallicity are not yet well constrained: $T_{\rm eff}=5100-5340K$ and $[{\rm Fe/H}]=0.20$ -- 0.45 \citep{arribas89,baliunas97,fuhrmann98,gonzalez98a,gimenez00,reid02,santos04,ecuvillon06}.  Many models show that 55\,Cnc is a sub-solar mass main-sequence star, whereas some other observations claim it to be a subgiant, based on the atmospheric CN enhancement \citep{greenstein68,taylor70,oinas74} and relatively low surface gravity, $\log{g}=4.10$ \citep{gonzalez97}.  However, using the greatly increased number of Fe~I and Fe~II lines, \citet{gonzalez98b} revised the surface gravity to $\log{g}=4.40$, which is too large for a normal subgiant star.  For our models, we have used the nine Lick/Hamilton spectra obtained by VF05.  The mean temperature and metallicity yielded by SME are $T_{\rm eff}=5253\pm44\,$K and $[{\rm Fe/H}]=+0.31\pm0.029\,$.  

The reported physical parameters of 55\,Cnc also show discrepancies.  Most of the analyses claim a model that is either very young and slightly more massive than the Sun ($\sim1.05M_\odot$) or very old and slightly less massive than the Sun ($\sim0.90M_\odot$).  Some of the spectroscopic analyses yield an age close to the Hubble time \citep{perrin77,cayrel87,gonzalez98a}.  FRS99 also predicts an old age $8^{-8}_{+7}\,$Gyr and a sub-solar mass $0.95M_\odot$.  Naively speaking, an extremely old age of 55\,Cnc seems unreasonable considering its anomalously metal-abundant atmosphere.  Chromospheric Ca~II activity of 55\,Cnc suggests instead a young age of $5\,$Gyr \citep{baliunas97}.  Using the high metallicity $+0.40\,$ and keeping the solar He-abundance, \citet{fuhrmann98} derived an age $\la5\,$Gyr.  It has even been suggested that 55\,Cnc might be a member of the Hyades Supercluster with an age upper limit of 2~Gyr \citep{eggen85,eggen92}.  However, various observational clues such as the measured stellar rotational period of $41.7\,$days, possibility of an extreme H-deficient atmosphere and the lack of detectable lithium, argue against extreme youth of the star \citep{gonzalez98b}.  The substantial size of the convective zone also supports a more evolved stellar model.

 Among the five sample stars we discuss here, 55\,Cnc has the least-constrained mass and age.  The observations by VF05 suggest it is a main-sequence star with $M=0.96^{-0.03}_{+0.05} M_\odot$ and $\FeH=+0.38$ (see Figure~\ref{55Cnc.fig}).  Due mainly to the slow main-sequence evolution of metal-rich stars with $M\la1.0M_\odot$, the derived age of the star is poorly constrained.  Although the age-PDF is indicative of an older age, nearly any age is possible within the range of 0 -- 14\,Gyr (Figure~\ref{five.fig}).  A choice of higher temperature and metallicity (e.g., $5336\,$K and $+0.40\,$, \citet{fuhrmann98}) would favor a younger age and a larger mass.  The situation can be understood as a result of the assumption for  He-abundance.  The scaling of He with respect to the other metals ($\delta Y/\delta Z$) is not well known, especially for these super-metal-rich stars.  Using the stellar evolutionary tracks with the high metallicity but keeping the solar He-abundance ($\delta Y/\delta Z=0$), \citet{fuhrmann98} showed that 55\,Cnc is located below the ZAMS line in the HR diagram.  If the H-depletion in the atmosphere is confirmed \citep{gonzalez98b}, He-fraction needs to be scaled up as $\delta Y/\delta Z=2.5\pm1.0$, as adopted by FRS99 and us.  Using a uniform prior distribution of $\delta Y/\delta Z$, our best fit  for 55\,Cnc occurs for our tracks with  $\delta Y/\delta Z=1.5$.    

The  model posterior temperature and metallicity, $5327\,$K and $+0.38\,$ are considerably higher than the values from the SPOCS catalog but similar to the observation by \citet{fuhrmann98}.  The derived surface gravity $\logg = 4.48^{-0.01}_{+0.05}$ is quite large and in agreement with VF05 and FRS99.

\subsubsection{$\rho$ Coronae Borealis (HD143761)} 
The AFOE (Advanced Fiber Optic Echelle spectrograph) team discovered a Jupiter mass planet in 39.8 days orbit around the G0 -- 2V star $\rho\,$CrB  \citep{noyes97a,noyes97b}.  It is the only star with sub-solar metallicity among the five stars discussed in this section.  The mean metallicities and effective temperature of the star yielded by one Keck/HIRES spectrum and three Lick/Hamilton spectra by VF05 are $T_{\rm eff}=5823\,$K and $[{\rm Fe/H}]=-0.14\,$.  Other spectroscopic analyses all identify the metal-poor atmosphere of $\rho\,$CrB, ranging from $\FeH=-0.19$ to $-0.32$ \citep{gratton96,kunzli97,fuhrmann98,gonzalez98a,gimenez00,henry00b,takeda01,santos04,ecuvillon06}.  

FRS99 yielded a model with an extremely old age, $14\pm2\,$Gyr, and $M=0.89\pm0.05M_\odot$.  Our analysis predicts a model that is slightly younger and more massive, $11.04^{-0.72}_{+0.88}\,$Gyr and $0.96\pm0.02M_\odot$.  This mass is consistent with the isochrone mass of $0.95M_\odot$ by \citet{santos04}, derived from the SARG spectrum.  The near turn-off main-sequence age has been confirmed by other isochrone / evolutionary track analyses; $10.2\pm1.7\,$Gyr \citep{fuhrmann98}; $12.1^{-2.0}_{+2.8}\,$Gyr \citep{nordstrom04}; $12.1\pm0.9\,$Gyr \citep{ng98}.  The paucity of heavy elements detected in the atmosphere is consistent with some of these very old ages, although metallicity is typically a poor age indicator \citep{saffe05}.  Other observational evidences support that $\rho\,$CrB is an evolved, near solar-mass star.  Relatively low chromospheric activity and slow rotational period ($\sim20\,$days) have been confirmed by \citet{noyes97a,henry00b}.  \citet{noyes97a} suggests that the high proper motion of $\rho\,$CrB out of the Galactic plane at $28\,\ks$ \citep{cayrel96} may indicate that the star was a member of the old disk population.  Our model with the slightly developed convective zone also supports that the star is at least near the end of the main-sequence stage.

The theoretically derived atmospheric parameters are consistent  with  the spectroscopic values by VF05, except that the model posterior surface gravity is significantly lower, $\log{g} = 4.18$.  This is consistent with the isochrone value of 4.14 by VF05, and other spectroscopic observations, 4.11 by \citet{gratton96} and 4.05 -- 4.19 by \citet{fuhrmann98}.

\begin{figure}
  \begin{center}
    \includegraphics[width=0.4\columnwidth]{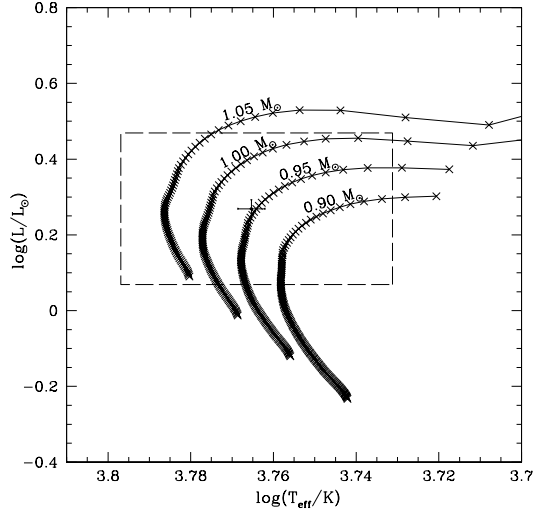}
		\caption{Theoretical HR diagram and stellar evolutionary tracks for $\rho\,$Coronae Borealis.  Tracks with our best-fit metallicity, $\FeH = +0.17$, are shown.    The tracks with $M=0.90M_\odot$ and $0.95M_\odot$ are truncated at the maximum age of the grids, $14\,$Gyr.  The derived model of $\rho\,$CrB  yields $M=0.96\pm0.02\Msol$ and $\tau = 11.04^{-0.72}_{+0.88}\,$Gyr.  \label{rhoCrB.fig}}
  \end{center}
\end{figure}

\subsection{Sample Models for Other Known Planetary Systems \label{additional_models}}

In addition to the stars previously modeled by FRS99, here we present the models for 5 planet-host stars with particularly interesting stellar or planetary properties that are worth detailed discussions.   Table~\ref{additional_models.tab} lists the spectroscopic parameters and the derived physical properties of these stars.

\subsubsection{HD209458}

HD209458 is the first extrasolar planetary system for which transit events were observed \citep{charbonneau00,henry00a}.  Accurate stellar modeling for  stars with  transiting planets is crucial for determining the radius and hence the interior structure of the planet.  The tightly constrained orbital inclination angles achievable for transiting planets can eliminate the factor of $\sin{i}$ from the observed radial velocity and yield an exact planetary mass ($\Mpl$).  From the photometric curve and the theoretically determined stellar radius, the radius  and therefore the density (and composition) of the planet can be found.

\citet{mazeh00} have created several stellar models using isochrones derived from the Geneva, Padova, Claret and Yale code.  By locating  the observed  $M_V$ and $\Teff$ and interpolating between the isochrones,   they determined $M=1.1\pm0.1\Msol$ and $R=1.2\pm0.1\Rsol$.  We have derived a similar model with better constraints, $M=1.13^{-0.02}_{+0.03}\Msol$ and $R=1.14^{-0.05}_{+0.06}\Rsol$.  Using the stellar parameters derived by \citet{mazeh00} and the observed inclination $i = 87^\circ.1 \pm 0^\circ.2$, \citet{charbonneau00} determined the planetary parameters, $\Mpl = 0.63 \MJ$ and $\Rpl = 1.27 \pm 0.02 \RJ$.  Using the theoretical isochrones by \citet{bertelli94}, \citet{allende99} determined a stellar radius which is closer to our result, $R = 1.15 \pm 0.08 \Rsol$.  \citet{henry00a} have adopted this model and derived a slightly larger planetary radius $\Rpl = 1.42 \pm 0.10 \RJ$ than the value by \citet{charbonneau00}.

\begin{deluxetable*}{lrrrcrrrrrrr}
	\tablecolumns{12}
	\tabletypesize{\scriptsize}
	\tablecaption{Sample Models for Stars with Known Planetary Companions \label{additional_models.tab}}
	\tablewidth{0pt}
	\tablehead{
		\colhead{}	&	\multicolumn{3}{c}{Observed Data}	&	\colhead{}	&	\multicolumn{6}{c}{Posterior Model Parameters} \\
		\cline{2-4}	\cline{6-11}	\\
		\colhead{Star}	&	\colhead{$\Teff$ [K]}	&	\colhead{[Fe/H]}	&
		\colhead{$\log{g}$ [cgs]}	& \colhead{}	&
		\colhead{$M_* [{\rm M}_\odot]$} & \colhead{Age [Gyr]} & 
		\colhead{$R_* [{\rm R}_\odot]$} &\colhead{$\Mce [{\rm M}_\odot]$} 
		&\colhead{$\Rce [{\rm R}_\odot]^a$} & \colhead{$\log{g}$ [cgs]} 
	}
	\startdata
	HD177830	&	4949	&	+0.33	&	4.03	&	&
	$>1.91$	&	3.24$^{-0.08}_{+0.56}$	&	
	2.95$^{-0.09}_{+0.15}$	&	0.945$^{-0.021}_{+0.114}$	&	$>1.29$	&
	3.91$\pm0.01$	\\
	
	HD209458	&	6099	&	+0.02	&	4.38	&	&
	1.13$^{-0.02}_{+0.03}$	&	2.44$^{-1.64}_{+1.32}$	&	
	1.14$^{-0.05}_{+0.06}$	&	0.007$\pm0.002$	&	0.24$^{-0.01}_{+0.02}$	&
	4.39$\pm0.04$	\\

	HD27442		&	4846	&	+0.29	&	3.78	&	&
	1.59$^{-0.14}_{+0.09}$	&	2.84$^{-0.36}_{+0.60}$	&
	3.43$^{-0.03}_{+0.11}$	&	1.014$^{-0.063}_{+0.048}$	&	$>1.87$	&
	3.56$^{-0.04}_{+0.05}$	\\

	HD38529		&	5697	&	+0.27	&	4.05	&	&
	1.48$\pm0.05$	&	3.28$^{-0.24}_{+0.36}$	&
	2.50$^{-0.06}_{+0.08}$	&	0.064$^{-0.009}_{+0.019}$	&	0.71$^{-0.01}_{+0.03}$&
	3.94$\pm0.02$	\\

	HD69830		&	5361	&	-0.08	&	4.46	&	&
	0.85$\pm0.01$	&	$> 12.04$	&
	0.90$\pm0.02$	&	0.038$^{-0.003}_{+0.002}$	&	0.29$\pm0.01$	&
	4.47$^{-0.01}_{+0.02}$	\\

	\enddata
	\tablecomments{Theoretical models for stars with known planetary companions discussed in $\S\,$\ref{additional_models}.  Adopted uncertainties for the observed parameters are, $44\,K$ for $\Teff$, $0.029\,$dex for [Fe/H] and $0.060\,{\rm cm\,s}^{-2}$ for $\log{g}$. }
	\tablenotetext{a}{Depth of the convective zone, measured from the outermost stellar surface}
\end{deluxetable*}

\subsubsection{HD69830}

The Doppler detection of three planets around the star HD69830 was recently reported by \citet{lovis06}.  It is the first triple-planet system consisting only of Neptune-mass planets: $\Mpl \sin{i} = 10.2\Mearth$ (planet~b), $11.7\Mearth$ (planet~c) and $18.1\Mearth$ (planet~d).  Prior to the planet detection, the system attracted a lot of interest owing to the large infrared excess observed by the Spitzer Space Telescope \citep{beichman05b}, indicating the presence of a massive asteroid belt within 1\,AU.   

Using the high resolution spectra obtained with HARPS at La Silla Observatory, \citet{lovis06} determined an effective temperature $\Teff = 5385\pm20\,$K and metallicity $\FeH = -0.05\pm0.02$.  The quoted values of $\Teff = 5361\pm44\,$K and $ \FeH =  -0.08 \pm 0.03$ by VF05 are in good agreement with these values.   The isochrone analysis using the theoretical evolution models by \citet{schaller92} and \citet{girardi00} yielded a stellar model of HD69830 with a mass $0.86\pm0.03\Msol$ and an age $\sim$4 -- 10\,Gyr.  Our calculation shows a similar model with a mass $0.85\pm0.01\Msol$ and an older  age $ > 12\,$Gyr.  \citet{lovis06} also performed numerical {\it N}-body simulations to test the dynamical stability of the system assuming coplanarity of the orbits.  They tested two inclination angles $i = 1^\circ$ and $90^\circ$, and in both cases the system remained stable for at least 1\,Gyr.  Note that long-term stability lasting  for as long as our derived  lower-limit age 12\,Gyr might favor smaller planetary masses, corresponding to a near edge-on view of the system ($i\sim 90^\circ$).  A more extensive stability analysis as well as a better-constrained stellar age are needed to provide tighter constraints on the orbital properties of the planets in this system.

\subsubsection{HD27442, HD38529 and HD177830}

The  SPOCS catalog contains 86 well-observed subgiants, and about 10 of these subgiants have detected planets.  Interestingly, three of them are observed to be super-metal-rich subgiants.  We find it worthwhile to present our theoretical models  for these subgiant stars because their physical  properties   are usually not  well-determined.  It is particularly difficult to  derive an accurate stellar mass for subgiants, since in the HR diagram theoretical subgiant tracks with different masses are much more closely separated  than  main-sequence tracks (see Figure~\ref{samplegrid.fig}).   On the other hand, derived ages of subgiants are relatively well-constrained,  because of the rapid cooling of the stellar atmosphere toward the red-giant phase.  

HD27442 \citep{butler01} is  one of the most evolved stars in our stellar sample, possibly already at the beginning of the red-giant phase.  It is a K\,2\,IV\,a star with a planet HD27442b with  minimum mass $\Mpl \sin{i} = 1.35 \MJ$ and  period $P = 423.8\,$days.  The observed parameters of the stars from the SPOCS catalog are $\Teff = 4846\,$K, $\FeH = +0.29$ and $\log{g} = 3.78$.  From spectroscopic observations, \citet{randich99} derived the atmospheric parameters $\Teff = 4749\,$K, $\FeH = +0.22$ and $\log{g} = 3.3$.  They further combined their observations with theoretical isochrones computed by \citet{bertelli94}, and determined the mass $1.2\pm0.1\Msol$ and age $\tau = 10\,$Gyr by interpolating  between two sets of isochrones, $\FeH = 0.00$ and $+0.40$.  We have derived a model that is younger and more massive than their model, $M = 1.48^{-0.08}_{+0.22} \Msol$ and $\tau = 2.84^{-0.36}_{+0.60}\,$Gyr.  

HD38529 is a G\,4\,IV subgiant star harboring two giant planets.  The inner  planet HD38529b ($\Mpl = 0.78 \MJ$, $P = 14.3\,$days) was  discovered by \citet{fischer01}.  A long-period residual trend  reported at the time was later confirmed to be another planet HD38529c, with a mass $\Mpl = 12.7\MJ$ and an orbital period $P=2174\,$days \citep{fischer03}.   The planetary masses are derived using the stellar model by \citet{allende99}: $M_* = 1.39 \Msol$, $R_* = 2.82 \Rsol$ and $\logg = 4.13$.  Interestingly, the large mass of HD38529c is close to the theoretical  deuterium burning limit and thus  suggests that it may be a sub-stellar companion rather than planetary.   We have derived the stellar mass $1.48\pm0.05\Msol$,  indicating a companion mass even larger than $12.7\MJ$.  

The K\,0\,IV star HD177830 is a highly evolved subgiant with very high metallicity, $\FeH = +0.33$.  A Jupiter-size planet HD177830b ($\Mpl \sin{i} = 1.22 \MJ$) with an orbital period of 391.6 days was discovered by \citet{vogt00}.   By interpolating the theoretical evolutionary tracks by \citet{fuhrmann97,fuhrmann98} to the observed $M_V$ and $B - V$, they estimated the stellar mass of $1.15\pm0.2\Msol$.  However, a   mass of a subgiant  star estimated by interpolating between evolutionary tracks can often be  inaccurate.  Moreover, modeling a star with such  high metallicity requires a grid of theoretical evolutionary tracks with a high metallicity resolution and a range of  helium fractions.  We have derived  a mass $M \ga 1.91\Msol$   and a radius $R = 2.62^{-0.05}_{+0.06} \Rsol$.   There is a small local maximum at $1.55\Msol$ in the mass-PDF with a $\sim10\%$ probability relative to the global maximum at $2.0\Msol$.  The age of HD177830 is well constrained: $\tau=3.24^{-0.08}_{+0.56}\,$Gyr.  HD177830 is a highly evolved super-metal-rich subgiant, located very closely to HD27442 in the HR diagram.  However, HD177830 is even more metal-rich  ($\FeH = +0.33$)  than HD27442, thus its mass needs to be at least comparable to that of HD27442 or even larger in order to reach the same evolutionary stage as HD27442 within a similar but slightly older age of $\sim3\,$Gyr.


\subsection{Parameter Correlations \label{correlations}}

In this section we explore possible correlations between the derived stellar parameters.  It should be remembered that there are many spurious trends that are artificially introduced during the stellar modeling process or because of observational selection effects.  Those artifacts need to be carefully removed to analyze any meaningful statistical correlations between the derived parameters.  

\begin{figure}
  \begin{center}
    \includegraphics[width=0.5\columnwidth]{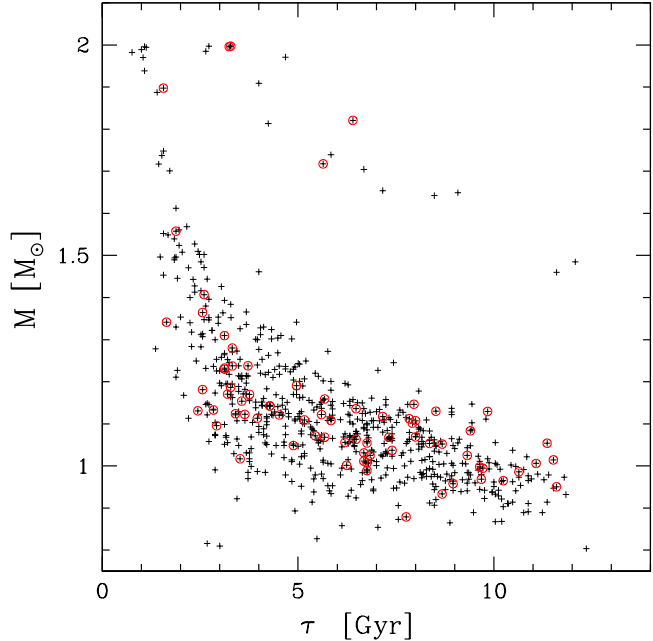}
		\caption{Age vs mass for 669 stars with well-defined ages.  The 74 planet-host stars are marked with red open circles.    \label{age_M.fig}}
  \end{center}
\end{figure}

\subsubsection{Age -- Mass Relations \label{age-M}}

Figure~\ref{age_M.fig} shows age -- mass relations for the SPOCS sample.  Only the 669 stars with well-defined mass and age are plotted, to remove the artificial accumulations at extreme ages (cf. $\S\,$\ref{bestestimate}).

Two major features are discernible in the figure: a large envelope of main-sequence dwarf stars extending from the upper-left (young, high-mass stars) to the lower-right corner of the figure (old, low-mass stars), and $\sim14\,$ subgiant stars, in the region $M>1.4M_\odot$ and $\tau>4\,$Gyr.  The shape of the large envelope is mostly determined by observational selection effects and our stellar modeling method: (i) The upper age limit for each stellar mass roughly corresponds to the main-sequence lifetime of the given stellar mass (e.g., $\sim10\,$Gyr for $1M_\odot$).  As the stellar mass increases, the age of main-sequence stars is typically better constrained.  Note that the wide range of metallicities creates a scatter in the main-sequence lifetime, since metal-poor stars typically evolve more rapidly (e.g., a $1\Msol$ star with only 10\% of solar metallicity leaves  the main sequence after $\sim7\,$Gyr).  (ii)  Each age has an approximate lower boundary with a critical mass below which well-defined age cannot be determined.  In general, for low-mass, unevolved stars near the ZAMS, only an upper bound of age can be derived, since these stars slowly evolve upward in the HR~diagram and do not leave the main-sequence track within $14\,$Gyr.  A few exceptions such as HD144253 and HD65583 in the region $M<0.85M_\odot$ and $\tau<4\,$Gyr have particularly low metallicities, [Fe/H] = -0.21 and -0.48, respectively.  Although the derived $1\sigma$ age uncertainties for these two stars are still large ($8.7\,$Gyr and $8.2\,$Gyr, respectively), these metal-poor stars experience a more rapid rise in luminosity during the main-sequence phase  which helps constrain the ages better than for slowly evolving metal-rich stars.   (iii) The narrow void on the left of the envelope is a lack of well-defined ages less than $\sim1\,$Gyr, mainly caused by the blue color cutoff in the SPOCS catalog ($B-V>0.5$) and exclusion of stars that are chromospherically very active.  

Similar trends are seen in the sample of planet-host stars.  Planet-host stars are dominant in the mass range $M=0.95-1.4M_\odot$.  Doppler radial-velocity observations become more challenging for stars with masses above this range because of the increasing atmospheric jitter.  Planetary companions are not common around low-mass later-type stars ($M \la 0.9\Msol$), although current spectroscopic surveys can achieve precise radial velocities for such low-mass stars.

\subsubsection{Age -- Metallicity Relations \label{age-FeH}}

\begin{figure}
  \begin{center}
    \includegraphics[width=0.5\columnwidth]{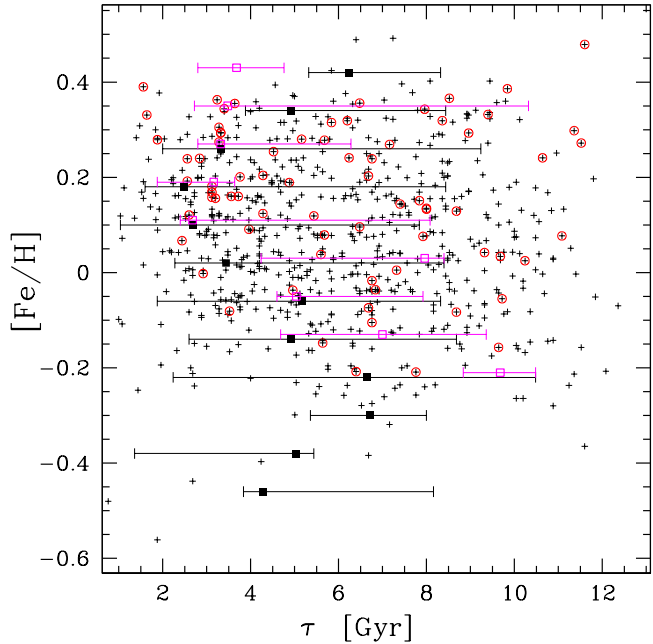}
		\caption{Derived well-defined age vs metallicity.  Here, red open circles represent  74 planet-host stars with well-defined ages.    An integrated age-PDF is calculated from all the age posterior PDFs for each metallicity bin, in order to estimate the age dispersion for given metallicity.   The mode value and 68\,\% credible interval of the stellar age in each bin are shown for the entire sample (solid square) and for the planet-host star sample (open square).    \label{age_FeH.fig}}
  \end{center}
\end{figure}

It has been confirmed by various spectroscopic observations that planet-host stars are on average more metal-rich than single field dwarfs \citep{gonzalez97,fuhrmann97,gonzalez98a,santos00,gonzalez01,santos01,santos03}.  One explanation posits that planets are more efficiently formed around stars with higher metallicity because of the higher fraction of solids available in the circumstellar disk \citep{ida04b}.  An alternative theory suggests that the observed high metallicities of planet-host stars are caused by late-stage accretion of gas-depleted material.  In this ``pollution'' or ``planet-accretion'' hypothesis, solid bodies (e.g., planetesimals or giant planet cores) migrate into the stellar atmosphere and enhance the stellar surface metallicity.  

 Theoretical calculations show that the degree of observable metal-enhancement is largely dependent on the size of the stellar convective zone (and therefore the effective temperature), but independent of the stellar age.  \citet{cody05} tested the effect of planet accretion on the subsequent stellar evolution by calculating various stellar evolution models with polluted stellar atmospheres.  They added metals to stellar convection zones at various arbitrary times up to $\sim 6\,$Gyr and showed that a polluted star will reach the same equilibrium state, regardless of whether the metal-rich material is accreted immediately or at a later time.  Their calculations suggest that polluted and thus metal-enhanced stellar atmospheres cannot be distinguished from intrinsically metal-rich stellar atmospheres in age -- metallicity relations.

Figure~\ref{age_FeH.fig} shows an age -- metallicity scatter plot for 669 stars with well-defined ages, including 74 planet-host stars.   It should be remembered that many planet-search programs preferentially select metal-rich stars to optimize planet detections, thus our sample is not expected to represent the true age -- metallicity relation for solar-neighborhood stars.   Another caveat is that  age is the most poorly-determined posterior parameter, thus a simple scatter plot for the mode ages and observed metallicities is not an accurate indicator of any possible age -- metallicity relation.    To see qualitatively the distributions of the derived ages and the observed metallicity, we have computed a sum of the normalized age-PDFs for each given metallicity.   In Figure~\ref{age_FeH.fig}, the mode  and associated $1\sigma$ uncertainties for the integrated age-PDF is presented for each metallicity bin.  Within the metallicity range  $\FeH=$-0.3 -- 0.4, the age is nearly uniformly scattered around the solar age ($\sim 4.5\,$Gyr) in each metallicity bin.   Note that the apparent mode-age -- metallicity relation is not a real trend but is an artifact of the age -- mass relation (Figure~\ref{age_M.fig}) and the mass -- metallicity relation (Figure~\ref{M_FeH.fig}).

While we do not find a significant age -- metallicity relation in the planet-host star sample, this may be largely due to the limitations of our analysis technique.  Several other studies of the age -- metallicity relation for planet-host stars also showed little or no variation in metallicity as a function of age \citep{saffe05,beichman05a,karatas05}.  The mean metallicity of the planet-host stars in the sample is higher by 0.12 than that of all the stars with well-defined ages, consistent with the analysis by \citet{fischer05}.  Interestingly, there are three planet-host stars with old ages ($\tau > 11\,$Gyr) and extremely high metallicity: HD30177 ([Fe/H]=+0.48), HD45350 ([Fe/H]=+0.30) and HD73526 ([Fe/H]=+0.27).  These three stars are all super-metal-rich late G-type stars with near solar mass, very similar to 55~Cnc.  The fact that most of the planet-host stars with very old ages ($\tau>10\,$Gyr) have super-solar metallicity largely comes from observational selection effects: (i) the frequency of planetary companions increases with stellar metallicity \citep{fischer05}, and (ii) stars with sub-solar metallicity evolve more rapidly.  Metal-poor F-,G- and K-stars most likely already left the main sequence before $10\,$Gyr and thus usually are excluded from planet-search programs.

\subsubsection{Metallicity -- Convective Zone Relations \label{Mce-FeH}}

\begin{figure}
  \begin{center}
    \includegraphics[width=0.65\columnwidth]{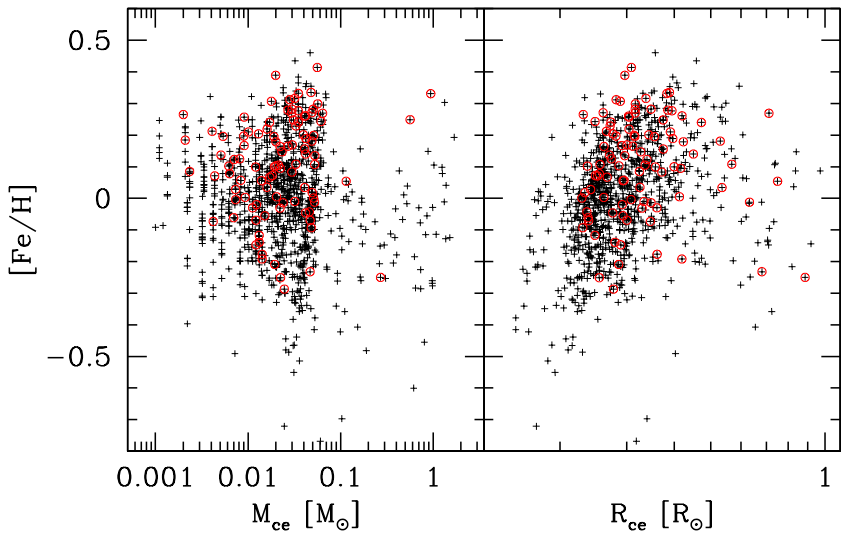}
		\caption{The mass and width of the convective zone vs metallicity.  Here $\Rce$ is defined as the distance between the stellar surface and the bottom of the convective zone.  Planet-host stars are marked with red circles.  The apparent deficit of stars with $\Mce\ga0.1\Msol$ comes from the relative shortage of K-type dwarfs in the SPOCS catalog and the steep decline of the convective envelope mass for stars earlier than F-type.  A possible weak correlation is present between $\Mce$ and maximum $\FeH$ in the range $\Mce=10^{-3}$ -- $10^{-1}\Msol$.     \label{Mce_FeH.fig}}
  \end{center}
\end{figure}
\begin{figure}
  \begin{center}
    \includegraphics[width=0.5\columnwidth]{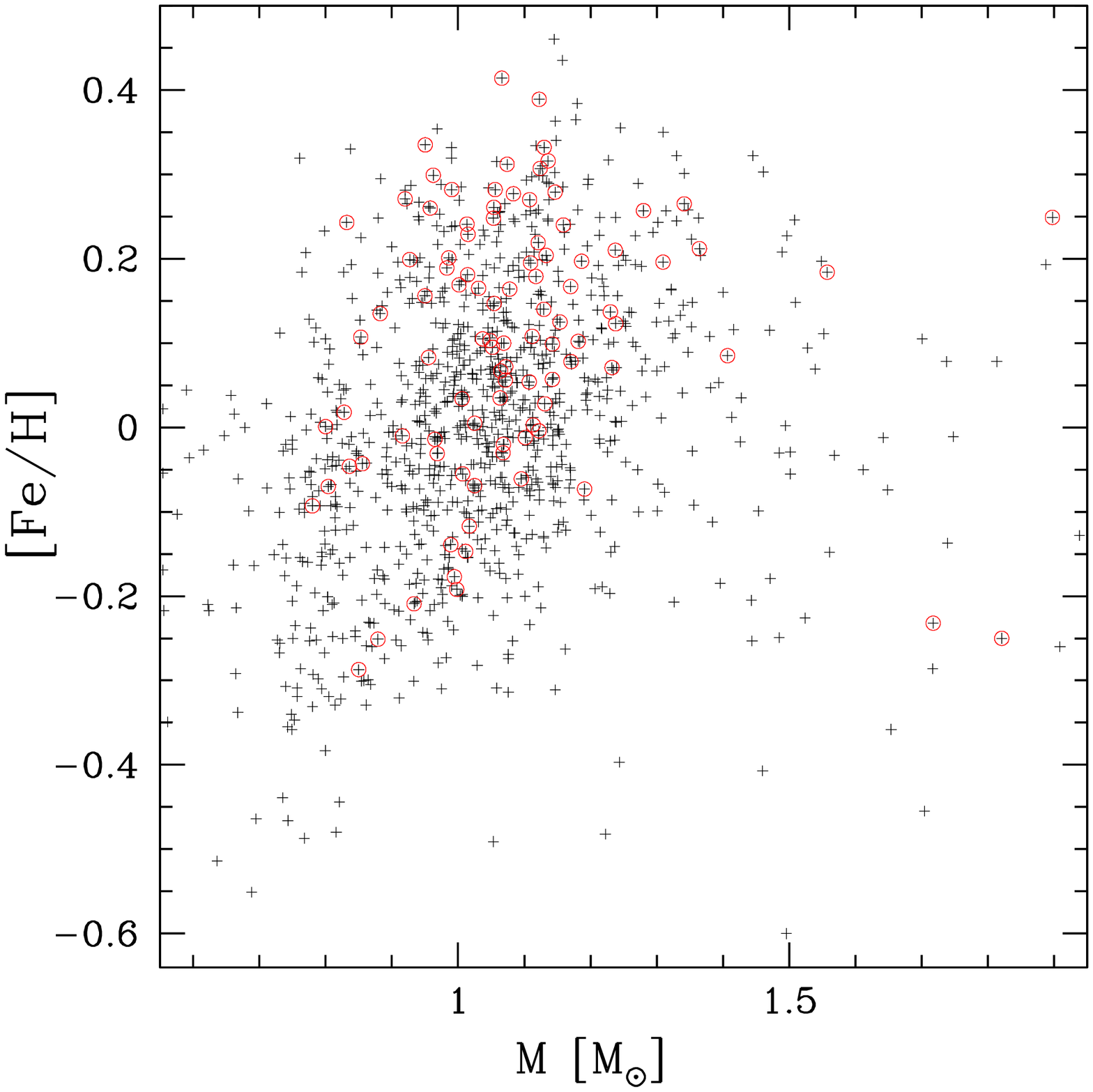}
		\caption{Stellar mass vs metallicity (left) and stellar mass vs convective envelope mass (right) for the entire sample.  The planet-host stars are marked with red circles.  Note that $\Mce$ starts to decrease faster for stars with masses above $1.0\Msol$.  \label{M_FeH.fig}}
  \end{center}
\end{figure}

The derived mass and depth of the convective zone are plotted against the observed metallicities in Figure~\ref{Mce_FeH.fig}.  The noticeable scarcity of stars with $M_{\rm ce} > 0.05M_\odot$ most likely comes from the relative shortage of K-type stars in the sample, corresponding to $T_{\rm eff} \la 5500K$ (see Figure~12 of VF05).  \citet{pinsonneault01} computed the mass of convective zone for stars with masses in the range 0.6 -- $1.3M_\odot$ and showed that $M_{\rm ce}$ is a sensitive function of the effective temperature.  For example, they showed that the convective envelopes of F-stars are more than 10 times less massive than that of K-stars (also see Figure~\ref{M_FeH.fig}).  The sharp decline of $\Mce$ for stars earlier than K-type, combined with the relative shortage of K-stars in the SPOCS catalog emphasizes the decrease in the population of stars with $\Mce \la 0.05 \Msol$ in Figure~\ref{Mce_FeH.fig}.

The distribution of $\Mce$ can be also used to test the metal-enrichment mechanism by planet accretion.   If planet accretion is a common phenomenon among planetary systems, we would expect an increase of maximum metallicities toward lower values of $\Mce$ because mixing of accreted material is less effective in a thinner convective envelope.  Thus, the enhanced metallicity is better preserved for stars with less massive convective envelope.  In particular, if planet accretion is responsible for the high metallicities of planet-host stars, then K-dwarfs and subgiant stars should show systematically lower metallicities, since the large convective envelopes of these stars completely dilute the accreted solids.  

Figure~\ref{Mce_FeH.fig} shows a weak {\em positive} correlation between the maximum $\FeH$ and $\Mce$, which is opposite of what is expected.  The most metal-rich planet-host stars in the sample are not F-type stars, but K- or G-dwarfs with larger convective zones: HD145675 (K0V, $\FeH=+0.41$), HD2039 (G2/3IV -- V, $\FeH=+0.39$) and HD30177 (G8V, $\FeH=+0.34$).  Note that F-type stars in the sample ($M=1.2-1.5\Msol$) are preferentially more metal-rich, because metal-poor F-stars have shorter main-sequence lifetimes (see Figure~\ref{M_FeH.fig}).  Despite this selection bias, the maximum metallicity and mean metallicities of the F-type stars in the sample are lower than those of the G- and K-type stars.  Furthermore, all three planet-host subgiants with well-defined ages are in fact super-metal-rich: HD27442 ($\FeH = +0.39$), HD38529 ($\FeH = +0.31$) and HD177830 ($\FeH = +0.36$).  This cannot be attributed to planet-pollution mechanism because subgiant stars develop large convective envelope in which accreted materials are well mixed with the deeper stellar interior.  Although a larger sample of K-dwarfs and subgiants would be more desirable, these results argue against significantly greater ($>0.1\,$dex) pollution of the convective envelope of stars with planets than stars without planets due to accretion of planets or planetesimals.

\subsubsection{Stellar Mass -- Planetary Mass Relations}

In spite of the chaotic nature of the formation and subsequent dynamical evolutions of planetary systems, it has been suggested that the distributions of planetary parameters may be mostly controlled by several key stellar parameters.  Extensive work has been done by \citet{ida04a,ida04b,ida05} in search for a deterministic theory of planet formation.  Planet-formation models can also predict subsequent orbital evolutions of planets such as migration or orbital decay by tidal dissipation, based on the initial configurations of the protoplanets.  Later evolution of planetary orbits may  also be characterized by the stellar properties.  Thus, the derived distributions of the planetary and stellar properties might indicate correlations that constrain certain formation and dynamical theories.

\begin{figure}
  \begin{center}
    \includegraphics[width=0.5\columnwidth]{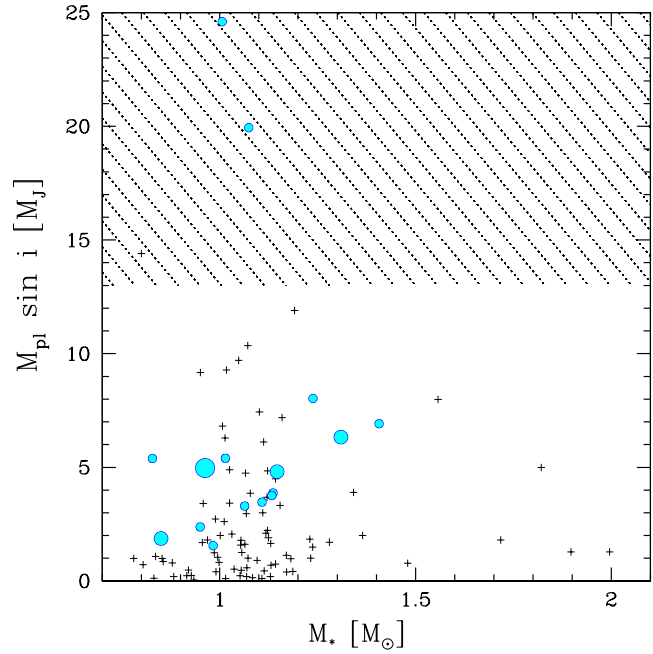}
		\caption{Stellar mass vs planetary mass.  For multiple-planet systems, the combined masses of all the planets are marked with circles.  The size of the circles increases in the order of double-, triple- and quadruple-planet systems.   The shaded region corresponds to the ``brown dwarf desert''.   The two systems above $\Mpl \sin{i} = 19 \MJ$, HD202206 and HD168443, are probably triple systems with a planet and a brown dwarf companion.  The companion of the star HD162020 with a minimum mass  $\Mpl \sin{i} =  14.4\MJ$ is also likely a brown dwarf.    The four stars with masses $>1.6\Msol$ are evolved subgiants.    \label{M_Mpl.fig}}
  \end{center}
\end{figure}

Figure~\ref{M_Mpl.fig} shows the relation between the observed planet mass ($\Mpl \sin{i}$) and the derived stellar mass ($\Mstar$).  The current planet formation theories based on planetesimal coagulation strongly depend on the properties of the circumstellar disk.  The growth of planetesimals is highly dependent on the surface density of the disk, thus stars with initially more massive disks are expected to produce larger planet masses.  However, the distribution of the disk masses inferred from infrared observations of T-tauri stars is not yet well-constrained.  The observations of dust in protoplanetary disks indicate that the total disk mass of a solar-type star can typically range from $10^{-4}$ to $10^{-1}\Msol$ \citep{beckwith96}.  Thus, accurate mass of protoplanetary disks cannot be determined as a function of stellar mass for the stars of our interest ($M=0.5$ -- $2.0\Msol$). 

  Using extensive numerical simulations, \citet{ida05} derived a positive correlation between the characteristic planetary mass and the stellar mass within the range 0.2 -- $1.5 \Msol$.  In Figure~\ref{M_Mpl.fig} the maximum planetary masses do increase in the same range of derived stellar masses.  However, the sample of only five systems is far too small to assert any correlation.   Also note that there is an intrinsically different population of companions above $\sim13\MJ$, corresponding to brown  dwarfs.  Spectroscopic observations show that there is a distinct absence of secondaries with masses $M_2\approx 0.01-0.08\Msol$ around solar-type primaries, so-called ``brown dwarf desert'' \citep{halbwachs00,grether06}.  This suggests a completely different formation channel for planets than for brown dwarfs.  Interestingly, the three systems with total companion mass beyond $13\MJ$ in Figure~\ref{M_Mpl.fig}, HD162020, HD168443 and HD202206, may be the rare candidates in which a K- or G-dwarf primary contains a brown dwarf companion.  The measured companion masses are, $\Mpl \sin{i} = 14.4\MJ$ (HD162020b), $17.5\MJ$ (HD202206b) and $16.9\MJ$ (HD168443c) \citep{udry02}.  HD202206 and HD168443 are triple systems, associated with another massive planet, HD202206c ($2.44\MJ$) and HD168443b ($7.7\MJ$).  It has not yet been confirmed whether these massive companions are sub-stellar or planetary.  HD162020b is particularly interesting, as it has a tight orbit similar to hot Jupiters ($P\sim8\,$days).  If it is confirmed to be a planet, that would significantly change the upper bound of the $\Mstar$ -- $\Mpl \sin{i}$ distribution.

Lastly, Figure~\ref{M_Mpl.fig} shows a lack of  massive planets around massive stars ($\Mstar>1.8\Msol$).  Doppler planet detection is intrinsically more challenging for stars earlier than F-type, and thus the sample in this region of parameter space is fairly incomplete.  Nevertheless, radial-velocity searches should favor detections of heavier planets.   Currently ongoing planet search programs targeting A -- F stars \citep{galland05a,galland05b} and G -- K giants \citep{sato05,hatzes05} will fill more samples in this region and may reveal interesting trends.

\subsubsection{Stellar Properties and Planetary Orbits}

Planetary orbits can still drastically evolve after the early phase of planet formation.  As a result, the distributions of orbital parameters reflect a mixture of various long-term dynamical histories as well as the initial configuration of planets, which may be partly obscured.  Thus, it is quite challenging to interpret any possible relations between orbital and stellar properties. 

\begin{figure}
  \begin{center}
    \includegraphics[width=0.5\columnwidth]{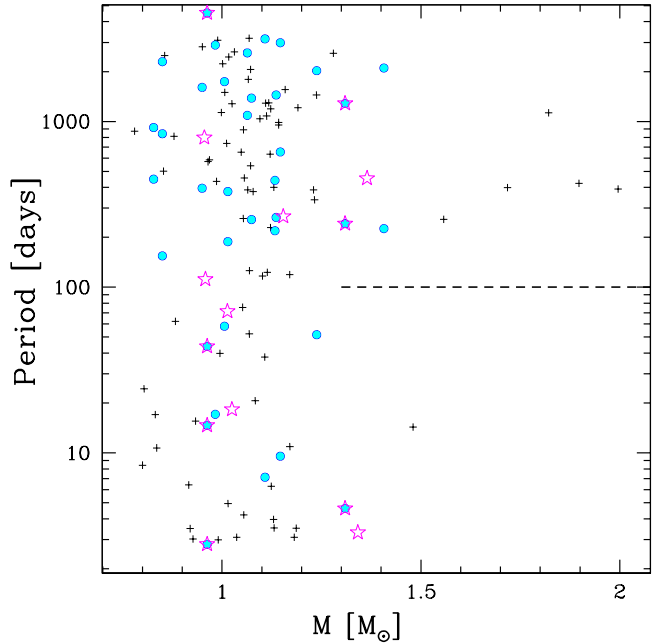}
		\caption{Stellar mass vs orbital period of planets.  Multiple-planet systems are denoted by solid circles.  Planets orbiting around a component of a stellar binary are denoted by open stars.  The period of $100\,$days corresponds to the boundary observed by \citet{udry03} below which no massive planet ($\Mpl>2\MJ$) is detected.  \label{M_P.fig}}
  \end{center}
\end{figure}

\citet{udry03} pointed out a lack of massive planets ($\Mpl \sin{i}>2\MJ$) with short periods ($P<100\,$days) in the observed sample of extrasolar planets.  Interestingly, a similar void may be observed in our stellar mass -- orbital period relation.  Figure~\ref{M_P.fig} shows a deficit of short-period planets around massive stars.  One possible explanation for the lack of high-mass close-in planets is that Type~II migration is less effective for more massive planets because of a larger gap-opening timescale \citep{trilling02}.  An alternative mechanism to eliminate short-period massive planets is the orbital decay due to tidal dissipation, although this is only   effective for orbital periods $\la 10\,$days \citep{rasio96a}.  Note that the five most massive stars above the void in Figure~\ref{M_P.fig} are all subgiants.  As a star evolves toward the red-giant phase, it develops a deep, massive convective envelope which more effectively dissipates the planet's orbital energy. \citet{rasio96a} demonstrated that if the star 51~Peg expands to about twice the Sun's radius, the orbital decay timescale drastically decreases by a factor of $\sim10^4$, and inevitably the planet will be swallowed in the stellar envelope.  Since the orbital decay time is also sensitively dependent on the initial orbital semimajor axis, planets at initially smaller distances have smaller chances of survival.  Since tidal dissipation is effective only within the range of a few stellar radii, it cannot explain the lack of planets with intermediate orbital periods ($P = 10\,$ -- $100\,$days) around massive stars.  Although the larger probability of fast Type~II migration during the formation and later tidal orbital decay may partly explain the lack of short-period planets around massive stars as well as the lack of short-period massive planets, more detections of planets around young massive stars and evolved subgiants are needed to further explore these ideas.

\begin{figure}
  \begin{center}
    \includegraphics[width=0.5\columnwidth]{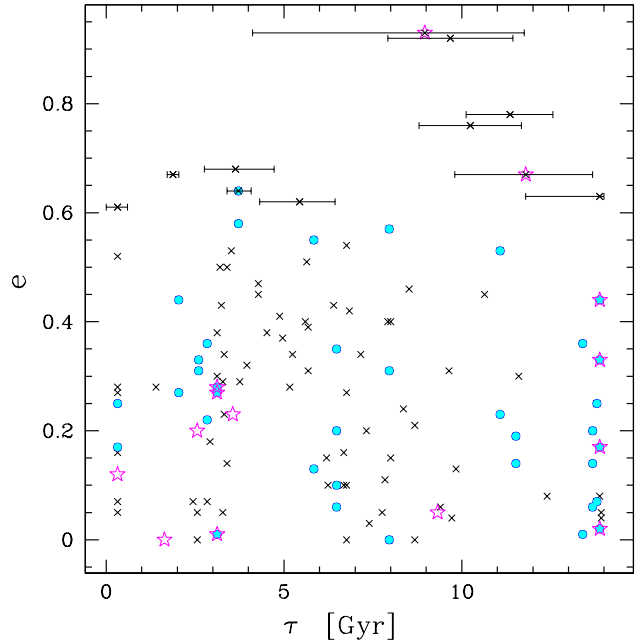}
		\caption{Derived stellar ages vs orbital eccentricities of planets.  Planets in multiple-planet systems are marked in solid circles.  The open stars represent planets orbiting around a component of a stellar binary.   Age uncertainties (68\,\% credible interval) are shown for planets with eccentricities larger than 0.6.  There is a lack of planets with very high eccentricity ($e>0.7$) and young age ($\tau < 4\,$Gyr).  \label{age_e.fig}}
  \end{center}
\end{figure}

Figure~\ref{age_e.fig} shows the orbital eccentricities of 120 extrasolar planets (99 systems) in the sample and the derived ages of the systems.  One of the striking orbital characteristics of extrasolar planets is their high orbital eccentricities.  As of May 2006, the mean eccentricity for 175 known extrasolar planets is 0.25, higher than that of Mercury (0.21) or Pluto (0.24) in our Solar System.  Planets formed in the standard scenario are expected to have a nearly circular orbits, thus the high eccentricities of extrasolar planets require later eccentricity perturbation mechanisms \citep[e.g.,][and references therein]{tremaine04}.  

Although there is no apparent age -- eccentricity correlation observed in Figure~\ref{age_e.fig}, there are a few signatures in the age -- e distribution that may indicate eccentricity evolution of the planets. (i) Most of the very eccentric planets ($e>0.6$) are in single-planet systems.  Two such planets are orbiting a component of a known stellar binary (HD80606 and 16~Cyg).  (ii) Most of the very high eccentricities ($e>0.6$) are observed in  old systems ($\tau \ga 5\,$Gyr).  (iii) Multiple-planet systems have a wide range of eccentricities with an upper bound around $e = 0.6$.  Systems with multiple giant planets of comparable masses can easily achieve high eccentricities through dynamical instabilities \citep{ford05} or crossing orbital resonances \citep{lee02,kley04}.  However, very high eccentricities can also lead to orbital crossings, and the planets ending up colliding or being ejected from the system.  While a larger sample is clearly needed, we find this particularly interesting since the planet-planet scattering model (with two planets on initially circular orbits) predicts a rapid decline in the frequency of eccentricities above $\sim 0.6$ and none above $\sim 0.8$ \citep{ford03}.  Secular perturbations from distant companions can also evolve planetary orbits into very high eccentricities.  When a planet orbits around a component of a wide binary system with a sufficiently large relative orbital inclination ($i\ga40^\circ$), the planet's orbit undergoes a long-term eccentricity oscillation with a maximum eccentricity up to almost unity \citep{kozai62,innanen97,takeda05}.  Theoretical models show that this secular perturbation (the ``Kozai mechanism'') can likely explain the high eccentricities of HD80606b \citep[$e=0.93$,][]{wu03} and 16~Cygb \citep[$e=0.67$,][]{holman97}.  Although the period of the eccentricity oscillation is  dependent on the semimajor axis and the eccentricity of the binary companion, it is typically on the order of Gyr.  Importantly,  the Kozai mechanism takes place dominantly in single-planet systems, since mutual interaction between multiple planets usually suppresses the secular perturbation.  The four single-planet systems in the region $\tau>9\,$Gyr in and $e>0.6$ in Figure~\ref{age_e.fig}, HD20782 \citep[$e=0.92$,][]{jones06}, HD45350 \citep[$e=0.78$,][]{marcy05a}, HD222582 \citep[$e=0.76$,][]{vogt00} and HD3651 \citep[$e=0.63$,][]{fischer03} are also good candidates where secular eccentricity oscillations could have taken place.  Since the Kozai mechanism can be effective in a fairly wide binary ($a > 1000\,$AU), it is still possible that a low-mass companion for these systems has remained undetected.  The results from ongoing searches for wide-orbit companions around extrasolar planetary systems \citep{mugrauer04,chauvin06} will help resolve these questions, particularly for old single-planet systems with very high orbital eccentricities.  It should also be brought to attention that in Figure~\ref{age_e.fig} there are at least a few single-planet systems in which large ($\sim 0.6$) eccentricities have been excited within a few Gyr.  Finding systems with very large eccentricities ($e\ga 0.8$) and young ages would be a challenge for the Kozai mechanism.  

\section{Summary}
	We have calculated theoretical stellar parameters ($M, \tau, R, \Mce, \Rce, \log{g}$, etc.) for 1074 stars from the SPOCS catalog.  Using Bayesian analysis, we have adopted an appropriate choice of the a priori stellar parameter distributions and computed posterior PDFs for each parameter.  We have provided several statistical summaries for each posterior PDF, such as the median, the mode, and various ranges of credible intervals, as well as flags indicating those cases for which a parameter is ``poorly-determined''.  The newly determined physical properties of the five sample stars proved to be consistent with the previous calculations by FRS99, but the high resolutions of our stellar evolution database provided stellar models with much smaller uncertainties ($\S\,$\ref{comparison}).  The complete list of the derived stellar parameters available in the electronic version of the paper are now ready to be used for various dynamical and formation studies of planetary systems.  Also, the computed database of stellar evolution tracks will continue to be a useful tool for modeling extrasolar planetary systems.  For example, the precise determinations of stellar radii applied to the large number of transit detections anticipated by OGLE or Kepler can provide important information for the interior structures of gas-giant planets.  

	The uniformly analyzed physical properties of the large stellar sample provided several interesting relations between stellar parameters.  The derived relations between convective zone mass and metallicity of known planetary systems seem to reject the atmospheric metal enrichment of planet-host stars caused by planet accretion.  We found no significant evidence of   stellar surface metallicity diluted in the large convective envelope of evolved stars, which was expected in the planet-pollution hypothesis.  

	The sample size of known planetary systems in the SPOCS catalog  still limits statistical analyses aiming to  identify correlations between  stellar  and planetary parameters.  The sample is mostly abundant with G-dwarfs due to their strong suitability for Doppler surveys.  A greater number of K- and F-dwarfs and evolved subgiants are needed to increase the leverage for  identifying correlations with stellar mass and radius.    Nevertheless, some of the features observed in the scattered plots in Figure~\ref{M_Mpl.fig}-\ref{age_e.fig} might provide clues about the history of formation and dynamical evolution in planetary systems:  

(i) There is a lack of close-in planets ($P<100\,$days) detected around subgiant stars with masses greater than $1.5\Msol$ (Figure~\ref{M_P.fig}).  This may be an effect of small number statistics.  However, it has been demonstrated by \citet{rasio96a} that the timescale for orbital decay due to tidal dissipation becomes progressively shorter as the planet-host star evolves into the subgiant phase.  The five long-period planets around subgiants may indicate that many of the short-period planets suffer orbital decay and will eventually be engulfed in the atmosphere of the giant stars.  Current planet searches around A -- F stars and subgiant stars will help constrain the possibilities of such dynamical process.

(ii) Many of the planets with extremely high eccentricities ($e>0.6$) are discovered in old systems (Figure~\ref{age_e.fig}).  Most of them are in systems with only a single known planet, and two of them are considered very likely to have been affected by secular perturbations induced by a stellar binary companion.  Secular orbital dynamics are likely to  be responsible for  such systems, since any impulsive perturbation can disrupt the planetary orbits in much shorter timescales.  It is of great interest to see whether the co-proper motion surveys in search for distant companions around known planetary systems \citep[e.g.,][]{mugrauer04,chauvin06} will confirm, or rule out the possibilities of stellar companions around these systems.

\acknowledgments 

This work was supported by NSF Grants AST-0206182 and AST-0507727 at Northwestern University. E.B.F. acknowledges the support of the Miller Institute for Basic Research.


\bibliographystyle{apj}                       

\end{document}